\documentclass[11pt]{article}

\usepackage{arxiv}

\usepackage[utf8]{inputenc}
\usepackage[T1]{fontenc}

\usepackage{geometry}
\usepackage{graphicx}
\usepackage{hyperref}
\usepackage{orcidlink}
\usepackage{textcomp}
\usepackage{lmodern}
\usepackage{makecell}
\usepackage{setspace}

\usepackage{natbib}
\usepackage{amsmath}

\newcommand{\cofirst}{\textsuperscript{†}}
\newcommand{\cosenior}{\textsuperscript{‡}}
\newcommand{\corrauthor}{\textsuperscript{*}}

\title{\textbf{Longitudinal MRI template of the baboon brain}\\from birth to adolescence}

\date{}

\begin{document}

\maketitle

\begin{center}
\begin{minipage}{0.9\textwidth}
\centering

Katherine L. Bryant\orcidlink{0000-0003-1045-4543}\cofirst\corrauthor$^{1,2}$ Arnaud Le Troter\orcidlink{0000-0002-7897-9672}\cofirst$^{3}$\\David Meunier\orcidlink{0000-0002-5812-6138}$^{3}$ Yannick Becker\orcidlink{0000-0002-9728-8316}$^{2,4}$ Scott A. Love\orcidlink{0000-0001-7416-9210}$^{5}$ Siham Bouziane\orcidlink{0009-0008-4042-6408}$^{1,2}$\\Kep Kee Loh\orcidlink{0000-0003-0650-224X}$^{1,3,6}$ Julien Sein\orcidlink{0000-0003-1767-5330}$^{1,3}$ Luc Renaud\orcidlink{0009-0003-5657-3406}$^{3}$\\Olivier Coulon\orcidlink{0000-0003-4752-1228}\cosenior$^{1,3}$ Adrien Meguerditchian\orcidlink{0000-0003-3754-6747}\corrauthor\cosenior$^{1,2,7}$

\end{minipage}
\end{center}


\vspace{1em}

\noindent
$^{1}$ Institute for Language, Cognition, and the Brain, CNRS, Aix-Marseille Université, Marseille, France\\
$^{2}$ Centre de Recherche en Psychologie et Neurosciences, UMR 7077, CNRS, Aix-Marseille Université, Marseille, France\\
$^{3}$ Institut de Neurosciences de la Timone, UMR 7289 CNRS, Aix-Marseille Université, Marseille, France\\
$^{4}$ Department of Neuropsychology, Max Planck Institute for Human Cognitive and Brain Sciences, Leipzig, Germany\\
$^{5}$ INRAE, CNRS, Université de Tours, PRC, 37380 Nouzilly, France\\
$^{6}$ Department of Psychology, National University of Singapore, Singapore\\
$^{7}$ Station de Primatologie, CNRS-CELPHEDIA UAR846, Rousset, France


\vspace{1em}

\noindent
\textsuperscript{†} These authors contributed equally to this work.\\
\textsuperscript{‡} These authors jointly supervised this work.\\
\textsuperscript{*} Corresponding authors: \\
\href{mailto:katherine.bryant@univ-amu.fr}{katherine.bryant@univ-amu.fr} \\
\href{mailto:adrien.meguerditchian@univ-amu.fr}{adrien.meguerditchian@univ-amu.fr}


\begin{abstract}
The baboon (Papio) is an invaluable resource within nonhuman primate research, having the
advantage of being a cercopithecoid (Old World monkey) with one of the largest brains among
non-hominid primates. In order to facilitate comparative developmental neuroscience
research, we present the $BABACOOL$ (BAby Brain Atlas COnstruction for Optimized Labeled
segmentation) approach for creating multi-modal developmental atlases, which we used to
produce $BaBa21$, a population-based longitudinal developmental baboon template. $BaBa21$
is a spatio-temporal template that consists of structural (T1- and T2-weighted) images and
tissue probability maps from a population of 21 baboons (Papio anubis) scanned at 4
timepoints beginning from about 2 weeks after birth and continuing to sexual maturity (5
years). Further, his study offers a fully automatic method for generating a template at any
intermediate age for future age-specific group studies. This resource is made available to
provide a normalization target for baboon data across the lifespan, including intermediate
timepoints, and moreover facilitate neuroimaging research in baboons, comparative research
with humans and nonhuman primate species for which developmental templates are available
(e.g., macaques).
\end{abstract}

\vspace{0.5em}

\section{Introduction}

As the field of non-human primate (NHP) neuroimaging expands, recent efforts have focused
on sharing resources and standardizing approaches to accelerate progress (\citet{Milham2018} \citet{Friedrich2021}). As new primate species are analyzed in a comparative neuroimaging framework, this not only advances our understanding of brain organization, but sheds light on the features that make human brains unique (\citet{Mars2014}), with clear
clinical implications (\citet{RogersFlattery2020}). Adding a longitudinal dimension to magnetic resonance (MR) templates offers further opportunities to understand the development of brain structures over time at both individual and population levels.

Population-based species-specific neuroimaging templates offer a normalization target that permits workers to register their primate acquisitions to a standard and thus effectively analyze their data in a validated anatomical space, and further, segment tissue types using established probabilistic priors. A previous baboon (Papio anubis) $T1w$ template, $Haiko89$ (\citet{Love2016}) is based on 89 individual adults. Here, we offer a developmental template generated from a cohort of 21 individual baboons (P. anubis) scanned at four time points, starting about 2 weeks after birth through adolescence.
Developmental templates, especially those based on longitudinal data from the same cohort, are labor- and resource-intensive, but offer a plethora of advantages to developmental neuroscience researchers (\citet{Cusack2018}). In pediatric studies, investigators are confronted with rapid brain changes due to postnatal maturation, necessitating age-specific templates. In recent years, developmental templates using longitudinal data have been generated for humans (e.g., \citet{Richards2016}) and macaques (e.g., \citet{Zhong2022}; \citet{Tan2024}). In addition to the utility of these templates for researchers with scans of neonates or juveniles, developmental templates in other NHP species open the door to investigations into the relationship between ontogeny and phylogeny. The most recent common ancestor of baboons (Papio) diverged from that of humans approximately 32 million years ago (\citet{Pozzi2014}), when crown Catarrhines (Old World anthropoids) split into the Hominoidea and Cercopithecoidea. Along with members of the Mandrillus genus, members of Papio possess the largest adult non-hominid brains among Primates in terms of absolute size (\citet{Isler2012}; \citet{Heldstab2018}; \citet{vanWoerden2012}); with P. anubis averaging approximately 160 cm3 (\citet{Meguerditchian2021}), over one-and-a-half times larger than rhesus macaques. Life history characteristics include a terrestrial lifestyle, omnivory, gestation of a single offspring, and a relatively long gestational period (6 months) during which the brain matures dramatically (\citet{Leigh2004}). Sexual maturity is reached by males and females between the 5th year and 6th years; however males take several additional years to reach full maturity in terms of size (\citet{Altmann2010}).
These life history characteristics, phylogenetic closeness to humans, and large brain size all make baboons valuable for understanding human brain evolution as well as human-specific neuropathologies. Specifically, baboons are a useful model for understanding the evolution of language (\citet{Fagot2019}; \citet{Meguerditchian2022}) and its cerebral hemispheric specialization
(\citet{Becker2022Symmetry}) social systems (\citet{Fischer2019}), and culture (\citet{Claidiere2018}). With regard to neuropathology, they are an established model for epilepsy (\citet{Szabo2012}) and stroke (\citet{Kwiecien2014}), and are more recently being used in studies of Parkinson’s disease (\citet{Arotcarena2020}) and age-related cognitive decline more broadly (\citet{Lizarraga2020}). Apart from the macaque (\citet{Zhong2022}) and marmoset (\citet{Sawiak2018}), few developmental templates exist for NHPs. Here we offer a population-based developmental multimodal ($T1w$ and $T2w$) baboon brain template with accompanying tissue probability segmentations in a common stereotactic space.

\section{Methods}
\label{sec:methods}

\subsection{Subjects}

All animal procedures were approved by the Ethical committee of neurosciences C2EA-71 (INT Marseille; APAFIS\#13553-201802151547729 v4), and were conducted at the Station de
Primatologie (SdP; Rousset-Sur-Arc, France) and the Centre IRM (INT Marseille, Centre
Européen en Imagerie Médicale) under agreement C130877 for conducting experiments on
vertebrate animals. All methods were performed in accordance with the relevant French law,
CNRS guidelines and the European Union regulations (Directive 2010/63/EU).
All subjects were free from developmental problems, neurological antecedents and brain
abnormalities. The facility housing baboons at SdP consists of outdoor enclosures with wood
and metal ethologically approved structures for enrichment. Feedings are four times a day and
consist of seeds, monkey pellets, fresh fruits and vegetables; water is available ad libitum.
The total number of scanned subjects was 31; of which 21 had a complete set of scans of high
quality and reached the last third longitudinal time point. These 21 were used in the
construction of the template. Due to segmentation failures in two cases and poor image quality
in one subject, 20 valid scans were available at both $ses-0$ and $ses-1$. As a result, the number
of usable images slightly varied across developmental stages (see Table \ref{tab:age_distribution}).

\subsection{Protocol}

Mothers and their infants were isolated from their social group the evening before the scan at
SdP for a medical check. Mother and newborn were transported together the following day to
the MRI center (Centre IRM, Institut Neurosciences de la Timone, Marseille, France). Upon
arrival at the MRI center, the mother of the focal subject was sedated with an intramuscular
injection of ketamine (3 mg/kg) and medetomidine (30 $\mu$g/kg). Neonates were then separated
from the mother and sedated using 4-7\% sevoflurane mixed with O2. Subjects over 5 months
old were sedated with 2-3\% sevoflurane. Tracheal intubation was performed using a
Sengstaken-Blakemore tube to assure artificial respiration in case of emergency. Tracheal
intubation was performed for steady controlled ventilation using an anesthetic ventilator
(Fabius MRI, Drager, Germany). Anesthesia was maintained with 3\% sevoflurane via a
calibrated vaporizer with a mixture of air 0.75 L/min and O2 0.1 L/min). O2 saturation and
cardiac rate were monitored throughout acquisitions with a 2 spO2 device placed on the lower
lip and a respiratory belt equipped with a mechanical sensor. End-tidal carbon dioxide,
halogen gas level, and volume/pressure was monitored and used to adjust ventilation rate (0.2
to 0.3 Hz) and end-tidal volume via an MRI respirator.
Each subject was placed in a ventral decubitus position in the scanner with the head secured
by foam positioners, cushions and velcro strips to maintain position and reduce potential
motion. Noise protection was attached around the ears. A forced-air warming blanket and hot
water bottles were used to maintain homeostatic temperature of the focal subject during all
procedures. Ophthalmic ointment (vitamin A) was applied to protect eyes from dryness. At the
end of the MRI session, subjects were monitored as they recovered from anesthesia.
Neonates and young baboons were returned to the mother in the cage after her recovery from
anesthesia. Mother and infant were monitored to make sure the mother responded to the
infant’s needs (breast feeding, etc). Focal subjects and mothers were returned to the SdP and
rejoined their social group the same or the following day.

\subsection{Imaging Parameters}

From September 2017 to March 2020, in vivo imaging was performed using a 3T clinical MRI
scanner (MAGNETOM Prisma, Siemens, Erlangen, Germany) equipped with 80 mT/m
gradients (XR 80/200 gradient system with slew rate 200 T/m/s), a 2-channel B1 transmit array (TimTX TrueForm). For timepoint 0, 1 and 2, two 11 cm receive-only loop coils with one under the head and another one around the face were used, and for timepoint 3, a 20-channel human head/neck coil was used. T1-weighted ($T1w$) images were acquired using a 3D-MPRAGE (\citet{Mugler1990}) sequence (0.4 mm isotropic, FOV=103x103x102.4 mm,
matrix=256×256, slices per slab=256, sagittal orientation, readout direction of inferior to
superior, phase oversampling=10\%, averages=3, TR=2500 ms, TE=3.01 ms, TI=900 ms, flip-
angle=8°, bandwidth=300 Hz/pixel, no fat suppression). T2-weighted (T2w) images were
acquired using the Sampling Perfection with Application optimized Contrast using different
angle Evolutions (SPACE) sequence (\citet{Mugler2000}), with the following parameters: 0.4
mm isotropic (0.5 for timepoint 3), FOV=154x115.5x102.4 mm, matrix=384×288, slice per
slab=256, sagittal orientation, readout direction I to S, phase oversampling=0\%, TR=3200 ms, TE=393 ms, bandwidth=566 Hz/pixel, no fat suppression, echo train length=790 ms and pre-scan normalization). The total acquisition time was 65 min for the structural imaging (35 min for $T1w$ and 30 min for $T2w$) out of a 2 hour session which included additional sequences
(BOLD, DWI, and QSM).

\subsection{Multimodal Template creation}

The $BaBa21$ template was constructed in a space compatible with the $Haiko89$ adult baboon
template (\citet{Love2016}): oriented in AC-PC, with a similar field of view, and a similar native spatial resolution - (0.4-0.5 mm)3 compared to $Haiko89’s$ (0.4 mm)3. This longitudinal version is comprised of acquisitions of the same individuals at four points in time, temporally distributed from newborn (2-4 weeks; “timepoint 0”), weaned (9 months; “timepoint 1”), juvenile (2 years; “timepoint 2”) and sexually mature (5 years; session "timepoint 3"). Twenty-one individuals of the entire dataset were selected for inclusion in the template (n=21; 12 males; Table \ref{tab:age_distribution}).
$T1w$ volumes were initially co-registered to the $Haiko89$ $T1w$ template (\citet{Love2016}),
downsampled to (0.6 mm)3, using the $antsRegistration$ function with the following parameters: rigid-only transformation, gradient step = 0.1, shrink factor (8x4x2), smoothing factor (3x2x1), and maximum iterations (1000x500x250), . Using these transformations, all volumes of the dataset were rigidly resliced into the $Haiko89$ space and left/right flipped in order to generate an approximately symmetrical template. More details are given in the sections below.

\begin{table}[h]
\centering
\caption{Age distribution of subjects across longitudinal timepoints.}
\label{tab:age_distribution}
\begin{tabular}{|c|c|c|c|c|}
\hline
\textbf{Time Point} & \textbf{Age Group} & \textbf{N (\# males)} & \textbf{Mean Age (days)} & \textbf{Mean Age (months)} \\ \hline
0 & Newborn & 20 (12) & 29.3$\pm$17.2 & 0.9$\pm$0.6 \\ \hline
1 & Weaned & 20 (12) & 275.8$\pm$35.1 & 9.2$\pm$1.2 \\ \hline
2 & Juvenile & 21 (12) & 745.2$\pm$21.7 & 24.8$\pm$0.7 \\ \hline
3 & Sexually Mature & 21 (12) & 1712.8$\pm$28.1 & 57.1$\pm$0.9 \\ \hline
\end{tabular}
\end{table}

\begin{figure}[h]
\centering
\includegraphics[width=\textwidth]{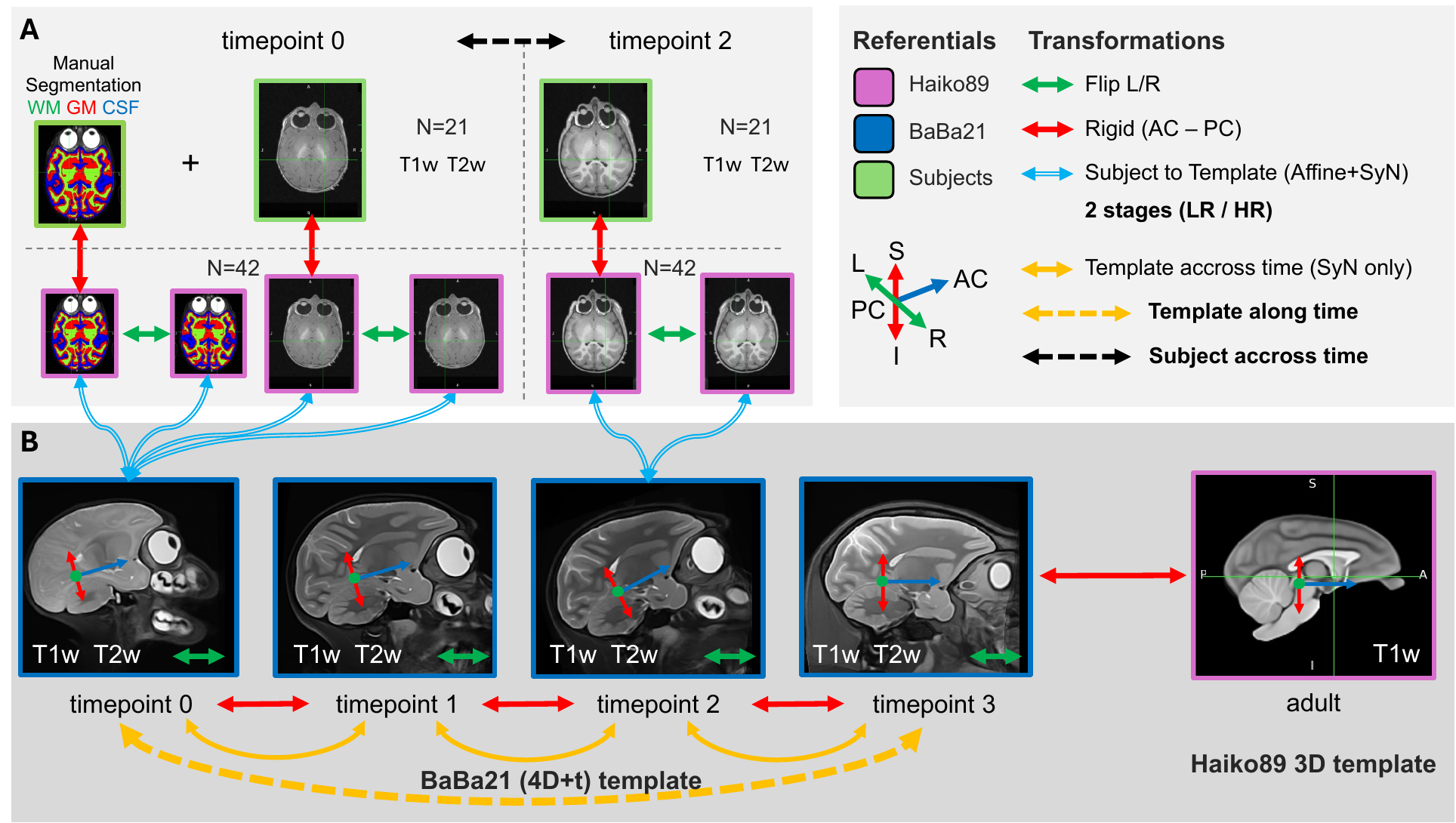}
\caption{Schematic of the $BaBa21$ template generation pipeline. First, population-averaged templates are generated for each timepoint (A); second, templates are normalized across timepoints via rigid AC-PC re-alignment, non-linear registration, and (4D+t) interpolation in one single field of deformation (B).}
\label{fig:baba21_pipeline}
\end{figure}

An iterative SyGN pipeline in ANTs (\citet{Avants2011}), using the $antsMultivariateTemplateConstruction2.sh$ script, was used in two steps to generate the multimodal template, pooling $T1w$ and $T2w$ contrasts ((Figure~\ref{fig:baba21_pipeline}.A)). In the first step, $T1w$ and $T2w$ images were downsampled to an isotropic voxel size of (0.6 $\mu$m)3, cropped to a matrix size of 1803 voxels in $Haiko89$ space, with maximum iterations set to 30x20x10. In the second step, the preprocessed $T1w$ and $T2w$ images were resampled to an isotropic voxel size (0.4 mm)3, cropped to a matrix size of 2703 in $Haiko89$ space, with maximum iterations set to 50x30x15. For both steps, the process consisted of four iterations with the following parameters: shrink factor 4x2x1, smoothing factor 2x1x0, and cross-correlation for pairwise registration. The output of the first step was interpolated to 0.4 mm isotropic resolution, serving as the input for the second step. Laplacian sharpening (\citet{XinWang2007}) was applied to the $T1w$ and $T2w$ templates at each iteration for both steps. For a detailed list of steps, \href{https://github.com/arnaudletroter/BABACOOL/blob/main/postprocessing/template_construction.md}{see \texttt{template\_construction.md}} in the BaBaCool github repository.

\begin{figure}[h]
\centering
\includegraphics[width=\textwidth]{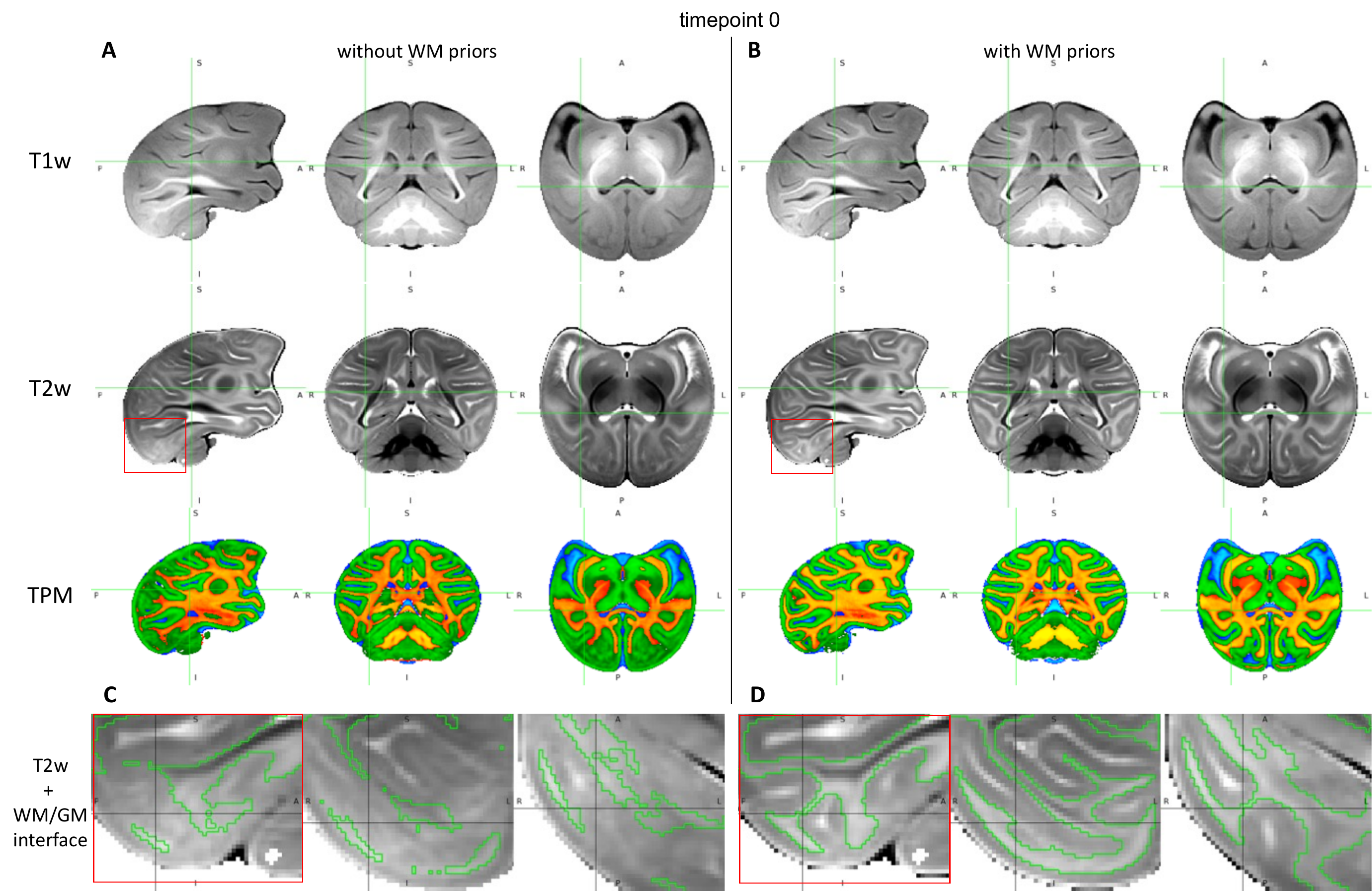}
\caption{Improvement of GM/WM contrasts and segmentation through
integration of mask of $WM$ during the SyGN process at timepoint 0. Whole-brain $BaBa21$
template without $WM$ priors (A) Whole-brain $BaBa21$ template with $WM$ priors (B) focus on
the temporal lobe and $WM/GM$ interface (C) focus on the temporal lobe and WM/GM
interface showing better spatial coherence (D)}
\label{fig:SupplFig1}
\end{figure}

For timepoint 0, the low $T1w$ contrast near the cortical ribbon poorly resolved the $WM/GM$ interface, thus, for this timepoint only, manually segmented $WM$ binary masks were integrated as a third contrast into template creation. SyGN registration was constrained by assigning weights (0.5 for $WM$ mask, 0.5 for $T1w$, 1.0 for $T2w$) during the maximization of the global similarity metric, permitting a clearer appearance of the cortical ribbon in the template space. A visual comparison of the template before and after the addition of the $WM$ mask prior demonstrates the improvement in contrast at the $WM/GM$ interface (Figure~\ref{fig:SupplFig1}).

\subsection{Alignment Across Timepoints}
\label{section_Alignment}

Templates from the four longitudinal sessions were rigidly aligned with AC-PC anatomical landmarks. The timepoint-(i) session was chosen as the target for the realigned timepoint-(i-1) session, with timepoint 3 being the initial target and $Haiko89$ the initial source. In the timepoint 3 session, three orthogonal planes based on the AC and PC landmarks were manually identified: the axial plane (xy, determined by [AC,PC] and [L,R] vectors), the inter-hemispheric sagittal plane (yz, determined by [AC,PC] and [S,I] vectors), and the coronal plane (xz, orthogonal to the others, determined by [L,R] and [S,I] vectors including the PC point; red, green and blue arrows in Figure~\ref{fig:baba21_pipeline}).

For each consecutive timepoint (timepoint 2, timepoint 1, timepoint 0), a multi-modal registration was determined based on $T1w$ and $T2w$ contrasts in two steps. First, a multi-stage transformation (rigid+affine+SyN) was performed to calculate precisely the location of the PC (the origin of the 3 planes) and the 3 planes. In a second step, the resulting backpropagated landmarks were used to estimate a single rigid transformation that can bring the PC and the three planes into the native space of the timepoints (timepoint 2, timepoint 1, timepoint 0) while preserving orthogonality of the 3 planes. The transformations were used to propagate segmentation masks from $Haiko89$ to timepoint 3, timepoint 3 to timepoint 2, timepoint 2 to timepoint 1, and timepoint 1 to timepoint 0. Next, the three planes were segmented into the four timepoints. The previous longitudinal registration scheme was repeated, using only rigid transformation based on a cross-correlation similarity metric estimated solely on the propagated segmentation masks. After this step, $BaBa21$ templates for the first three timepoints were reoriented across the four timepoints. Finally, the multi-modal templates were left/right flipped and averaged to generate a temporary symmetrical template, which was used as the final target for a last rigid registration step. This step produced an exact symmetrical template by averaging both the co-registered template and its flipped version.

\subsection{Tissue Probability Maps and Binary Brain Mask Generation}

Tissue probability maps ($TPM$) for white matter ($WM$), gray matter ($GM$), and cerebrospinal fluid ($CSF$) were generated in the following manner. After automatic segmentation with Macapype (macatools.github.io/macapype), $WM$ segmentations were manually corrected in subject spaces by an expert in anatomy ($SB$) for timepoint 0, timepoint 1, and timepoint 2. Timepoint 3 did not require manual corrections due to the image quality being similar to adult scans. Timepoint 0 required more substantive corrections in all brain regions (especially the cerebellum) than other timepoints, which was achieved by utilizing $T2w$ contrasts, which offered better visibility of the $WM/GM$ interface than $T1w$ images alone (Figure~\ref{fig:SupplFig1}). All final segmentations of $GM$, $WM$, and $CSF$ tissues were deformed by applying all geometric transformations rigid to $Haiko89$ and rigid + affine + nonlinear to $BaBa21$ spaces and averaged to build the new tissue probability maps.

Binary thresholding (values greater than 0.2) was applied to the TPMs for $GM$ and $CSF$ . The resulting binary masks were combined, followed by a hole-filling operation (26-connectivity). Morphological erosion using a 3D kernel (3x3x3 voxels) was then performed to generate the final binary brain mask (BM). The TPMs for $GM$ and $CSF$ were generated by masking them with the BM. The $WM$ TPM was subsequently regenerated from the binary BM by subtracting the corrected TPM values for $GM$ and $CSF$ to ensure that the sum of the $GM$, $WM$, and $CSF$ probability values in each voxel equals 1. 

\subsection{Histogram-Based Normalization for $T1w/T2w$ Intensity Transformation}
\label{sec:histo}

The Laplacian filter applied during the $SyGN$ process, after averaging for each iteration, enhances the tissue contrasts and improves the resolution of anatomical details. This permits improved segmentation and registration between subject and template, but also contributes to increasing the differences in $T1w$ and $T2w$ value intensity ranges across timepoints. To reduce these variations, $T1w$ and $T2w$ maps were normalized using multiclass histogram-based approach (\citet{Sun2015}), using the following equation~\ref{eq:normalization}. :

\begin{equation}\label{eq:normalization}
\begin{aligned}
T1w_{\text{norm}}(x, y, z) &= a \cdot T1w(x, y, z) + b, \\
T2w_{\text{norm}}(x, y, z) &= c \cdot T2w(x, y, z) + d.
\end{aligned}
\end{equation}

where $(a,b)$ and $(c,d)$ are respectively variables of a system of linear equations ~\ref{eq:ab} and ~\ref{eq:cd},
respectively for normalization of $T1w$ and $T2w$ maps as described below:

\begin{minipage}{0.48\textwidth}
\begin{equation}\label{eq:ab}
\begin{aligned}
70 &= a \cdot T1w(WM)_{90\text{th}} + b \\
30 &= a \cdot T1w(GM)_{10\text{th}} + b
\end{aligned}
\end{equation}
\end{minipage}
\hfill
\begin{minipage}{0.48\textwidth}
\begin{equation}\label{eq:cd}
\begin{aligned}
30 &= c \cdot T2w(WM)_{10\text{th}} + d \\
70 &= c \cdot T2w(GM)_{90\text{th}} + d
\end{aligned}
\end{equation}
\end{minipage}

Note that Laplacian sharpening is optional for this step of template generation. Also note that prior to this intensity normalization, a $N4$ bias correction is applied on templates in order to remove any intensity bias effects. It is also possible to correct for the B1 bias in all images prior to template construction. Tests showed that the resulting template is equivalent to the method presented here. Nevertheless, the pipeline offers the possibility to use both strategies (\href{https://github.com/arnaudletroter/BABACOOL/blob/main/postprocessing/bias_correction.md}{bias correction} before or after template construction).

\subsection{Generation of Intermediate Timepoint Templates}

To interpolate intermediate timepoint templates, a multimodal $SyNonly$ registration was
applied to the volumes that were previously rigidly aligned (see Section~\ref{section_Alignment}). Once the multi-modal $SyNonly$ registration, based on $T1w$ and $T2w$ contrasts, converged across two
successive timepoints, warp and inverse warp deformation fields were generated. Then, each
$BaBa21$ template modality $(T1w,T2w,TPM)$ was deformed over time along the series of
diffeomorphisms connecting them, creating a linear interpolation along the $SyN$ geodesic path
(\citet{Avants2008} \citet{Fu2017}) between timepoint 0 and timepoint 1. Next, each $BaBa21$
template modality $(T1w, T2w, TPM)$ was deformed over time along the series of
diffeomorphisms connecting them. Intermediate volumes were generated by applying
fractional $SyN$ transformations along the geodesic path. For each image pair, forward and
inverse diffeomorphic transformations were estimated using the Symmetric Normalization
$(SyN)$ algorithm implemented in ANTs (\citet{Avants2008}). Let $\phi_1$ denote the forward
transformation from image $I$ to image $J$, and $\phi_2$ the corresponding inverse transformation. The
warped images $\phi_1(I)$ and $\phi_2(J)$ were combined to generate intermediate volumes according to

\begin{equation}
\label{eq:Kt}
K(t) = \frac{t \phi_1(I) + (1 - t) \phi_2(J)}{2}
\end{equation}

where $t \in [0,1]$ weights the relative contribution of the forward-warped source image and the
inverse-warped target image.

\begin{figure}[h]
\centering
\begin{tabular}{cc}
\includegraphics[width=200pt]{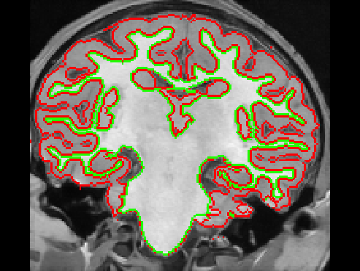} &
\includegraphics[width=200pt]{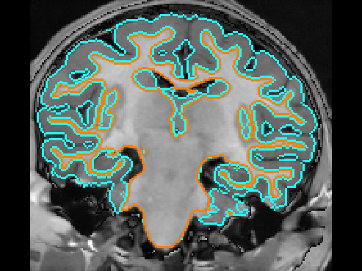} \\
\makecell[l]{(a) $Ozone_{ses1}$ segmentation, \\ $V_{WM}(Ozone_{ses-1})=35676$ (green)} &
\makecell[l]{(b) $Ozone_{ses1}$ segmentation warped onto $Ozone_{ses0}$, \\ $V_{WM}(Ozone_{ses-0})=33971$ (orange)}
\end{tabular}
\caption{WM segmentation and estimated $WM$ volume change between $Ozone_{ses-0}$ and $Ozone_{ses-1}$.}
\label{fig:comparison_volumes}
\end{figure}

To further investigate $WM$ consistency across timepoints and qualitatively assess the behavior of the interpolation function, $WM$ TPMs were thresholded at 0.5 to generate binary masks, from which surfaces were subsequently extracted for each original and interpolated timepoint.
This surface-based representation enables a direct visual examination of the spatial continuity induced by the interpolation, allowing us to evaluate whether intermediate shapes evolve smoothly and remain anatomically consistent with the original timepoints. To examine the accuracy of the intermediate timepoint templates, we compared the $BaBa21$ template (broken down into 11 timepoints between $ses-0$ and $ses-1$) to a single subject (Ozone) which had not been included in our template generation due to the fact that it was scanned at a timepoint intermediate to $ses-0$ and $ses-1$. This comparison allows us to determine whether the intermediate interpolated anatomy reasonably fits the measurements for $Ozone$ at the actual intermediate timepoint. We quantified this fit by estimating the amount of volumetric changes between a given timepoint and $ses-1$ of $Ozone$. $Ozone$ was scanned at 165 days (between $ses-0$ and $ses-1$ timepoints for $BaBa21$) and 265 days (within the range of $ses1$ for $BaBa21$). To avoid confusion with the $Baba21$ timepoints, we will call these two time points $Ozone_{ses-0}$ and $Ozone_{ses-1}$ in the text below. These scans were compared to 11 models from $BaBa21$: $ses\-0$ (29 days), $ses\-1$ (275 days), and 9 models of interpolated intermediate timepoints between those two timepoints. Ozone’s session 1 scan was manually segmented with the same methodology that was applied to the $BaBa21$ subjects. $WM$ and $GM$ maps were then non-linearly registered (using rigid+affine+SyN transforms) to $Ozone_{ses-0}$. The $WM$ volume change between $Ozone_{ses-0}$ and $Ozone_{ses-1}$ was then estimated (Figure~\ref{fig:comparison_volumes}).

For each $BaBa21$ timepoint, we compared the volume of $WM$ between a given timepoint and the $Baba21_{ses-1}$. We can then assess which of the $BaBa21$ timepoints has the most similar $WM$ volumetric change to the change observed in Ozone between $Ozone_{ses-0}$ and $Ozone_{ses-1}$. Volume changes are then compared to the reference ($Ozone_{ses-0}$ to $Ozone_{ses-1}$) changes by computing a volume ratio similarity $V_{RS}$ as follows: $WM$ volumes are computed as the sum of values in the $WM$ $TPM$ of each image, and the similarity is :

\begin{align*}
V_{RS}(t) &=  \frac{V_{WM}(Ozone_{ses-1} )}{V_{WM}(Ozone_{ses-0} ) }-\frac{V_{WM}(BaBa21_{ses-1} )}{V_{WM}(BaBa21_{ses-t} )}
\end{align*}

\section{Results}
\label{sec:results}

\subsection{Templates and Tissue Probability Maps}

\begin{figure}[h]
\centering
\includegraphics[width=\textwidth]{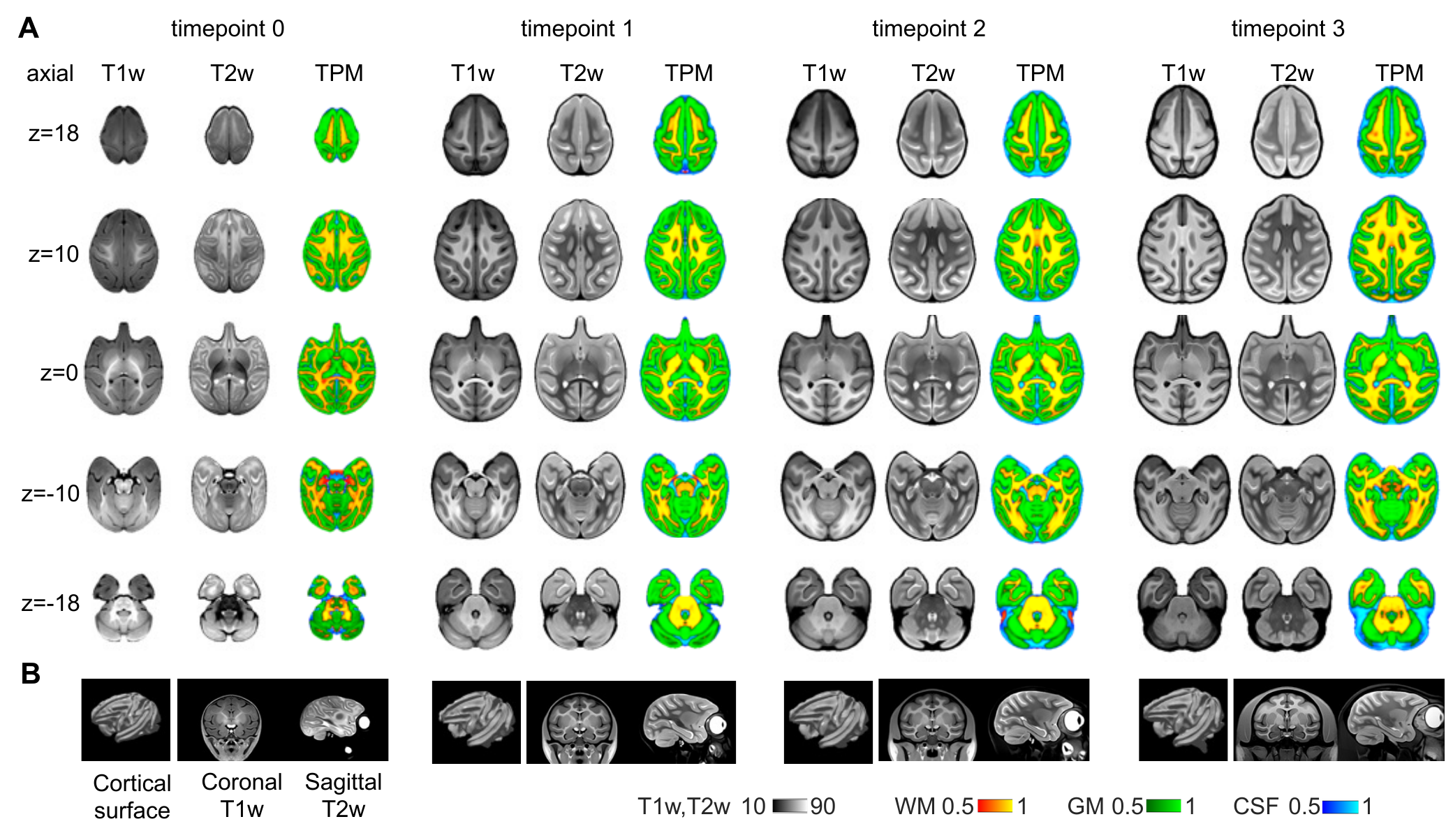}
\caption{$BaBa21$ (4D+t) template with (A) representative axial slices from $T1w$, $T2w$, $TPM$
volumes by column at 5 positions along the z axis (z=-18, z=-10, z=0, z=10, z=18) and (B) 3D
cortical surface, coronal $T1w$ and sagittal $T2w$ views.}
\label{fig:figure2}
\end{figure}

Multimodal population-averaged templates integrating $T1w$ and $T2w$ images were created for
all four timepoints as described in section ~\ref{sec:methods} (See Figure\ref{fig:figure2}). As described in section 2.6, the classification of tissue types into gray matter, white matter, and cerebrospinal fluid was generated as part of the segmentation within the $BABACOOL$ pipeline (Figure\ref{fig:figure2}; see Supplementary Figure\ref{fig:SupplFig3} for manual tissue segmentation for an individual subject across all timepoints). TPMs were precise enough to resolve even narrow $WM/GM$
boundaries in the extreme capsule at timepoint 0 (Figure\ref{fig:SupplFig1}).

\begin{figure}[h]
\centering
\includegraphics[width=\textwidth]{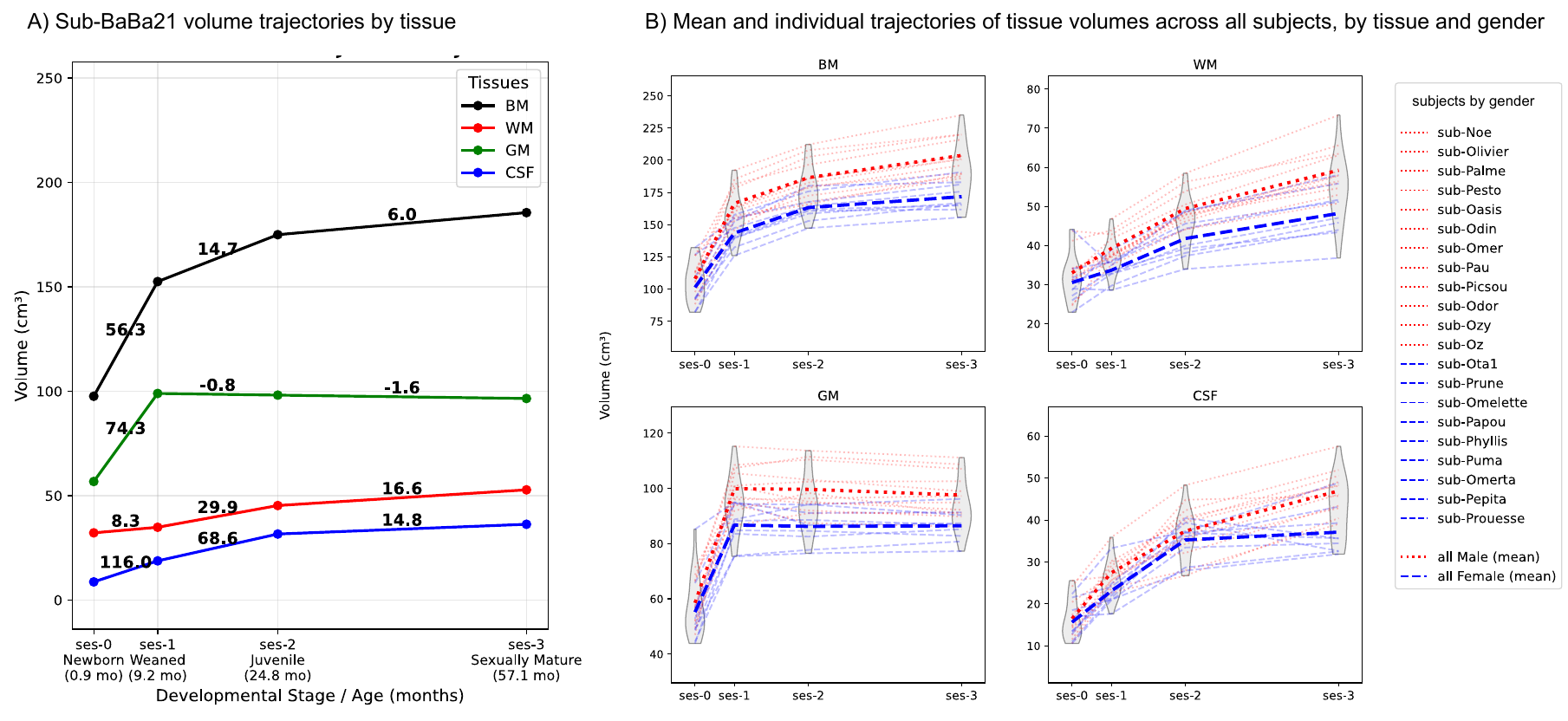}
\caption{Tissue volume trajectories across timepoints at the template level (A) together with the
same trajectories per subject and split by sex (B).}
\label{fig:figure3}
\end{figure}

TPMs were optimized using a histogram-based normalization approach (described in section~\ref{sec:histo}. Before normalization, inter-timepoint variability in $T1w$ and $T2w$ values were apparent (Supplementary Figure\ref{fig:SupplFig4}.A), while after, tissue intensity values were centered within 30-70\% of the
normalized maximum intensity (Supplementary Figure\ref{fig:SupplFig4}.B). In order to properly describe the demographics
of our group, we provide in Figure \ref{fig:figure3} the tissue volume trajectories across timepoints at the
template level (Figure\ref{fig:figure3}.A) together with the same trajectories per subject and split by gender (Figure\ref{fig:figure3}.B)

\subsection{Intermediate timepoint templates}

Intermediate timepoints for all template modalities ($T1w$, $T2w$, $TPM$) were linearly interpolated between timepoint 0 and timepoint 1 (Figure\ref{fig:figure4}). Normalized $T1w$, $T2w$, and $TPM$ volumes are
represented in rows (on a single axial slice at z = 0), for a set of intermediate interpolated
timepoints K at time t = [0,0.2,0.4,0.6,0.8,1], shown in columns (Supplementary Figure\ref{fig:figure4}). The last two rows
show intermediate deformation fields, weighted by the interpolation factor t, are visualized through the logarithm of Jacobian determinant and the global deformation field from timepoint 0 to timepoint 1 as a vector field (Supplementary Figure\ref{fig:figure4}) and the full forward deformation fields between successive timepoints (Supplementary Figure\ref{fig:SupplFig5})

\begin{figure}[h]
\centering
\includegraphics[width=\textwidth]{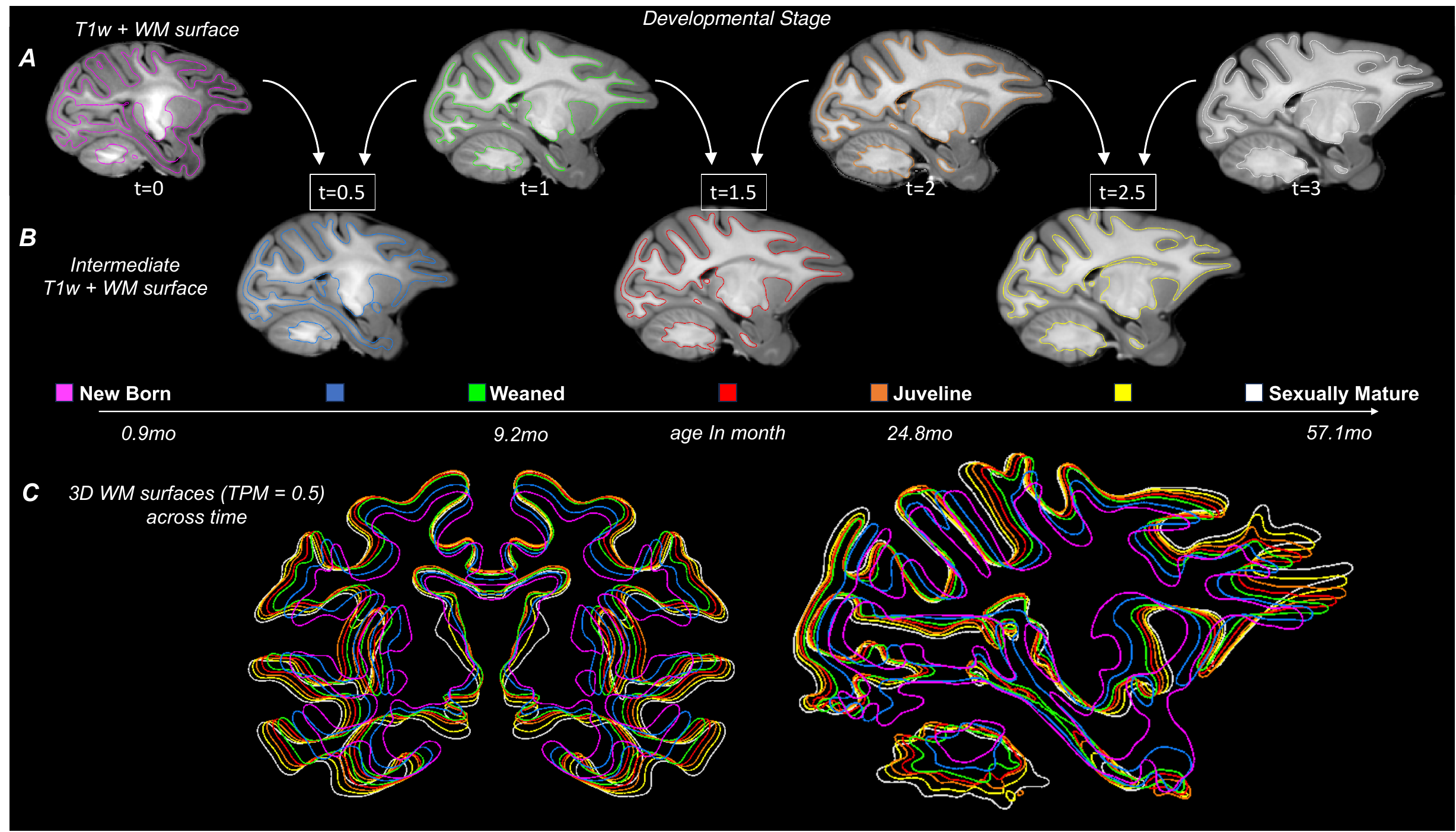}
\caption{(A) Sagittal views for the four original T1w/WM $BaBa21$ templates (timepoints
t=0,t=1,t=2,t=3) and (B) Three intermediate template generated by the geodesic path across
time (timepoints t=0.5,t=1.5,t=2.5), (C) Coronal and Sagittal views of 3D White matter (WM)
surfaces extracted from thresholded tissue probability maps (TPMs) (threshold = 0.5, row C)
across time.}
\label{fig:figure5}
\end{figure}

Comparing the interpolated timepoints between ses-0 and ses-1 with values from an individual subject scanned between those timepoints (175 days) showed that the interpolated estimate was close to the measured value (Supplementary Figure\ref{fig:SupplFig6}). The resulting $WM$ surfaces derived from thresholded TPMs are shown in Figure\ref{fig:figure5}. Sagittal views of the four original templates (t = 0, 1, 2, 3) together with the intermediate templates generated along the geodesic trajectory (t = 0.5, 1.5, 2.5) demonstrate a continuous temporal progression of anatomical shape. The $WM$ surfaces exhibit smooth spatial transitions between consecutive timepoints, without visible discontinuities or abrupt geometric changes, supporting the stability of the interpolation process. Overall, these qualitative observations
confirm the anatomical plausibility and temporal coherence of the interpolated templates.

\section{Discussion}
\label{sec:discussion}

\subsection{Summary}

$BaBa21$ is a longitudinal developmental baboon brain template, complementing and building on a previously released adult baboon brain template, $Haiko89$ (\citet{Love2016}), which uniquely integrates $T1w$ and $T2w$ contrasts. This template includes 4 timepoints from neonate to sexual maturity, providing a longitudinal averaged template designed to measure tissue evolution over development. This is an essential resource for brain development studies that rely on quantitative morphometric analyses performed directly from tissue segmentation, and further, facilitates the multi-species comparisons necessary for investigating the relationship between brain development and evolution.

Like $Haiko89$, $BaBa21$ is a symmetric template, which facilitates analysis of lateralization through voxel-based morphometric approaches. In humans, the link between tissue volume differences between hemispheres and function is well-established, with left-lateralized white matter volumes observed in human language areas (\citet{Pujol2002}), and correlations between language lateralization and relative gray matter volumes (\citet{Josse2009}). In recent years, the baboon has been explored as a model for understanding structural and functional lateralization. Baboons show human-like lateralized motor behaviors such as handedness for manipulative actions (\citet{Vauclair2005}, \citet{Molesti2016}) associated with hemispheric specialization of the central sulcus (\citet{Margiotoudi2019}), but also patterns of hand preference associated specifically with gestural communication (\citet{Meguerditchian2006}; \citet{Meguerditchian2011}). Baboons also have structural brain lateralization in areas homologous to human language areas, including the surface area of the planum temporale in adults (\citet{Marie2018}) and newborns (\citet{Becker2021}) as well as the Broca’s area homolog (\citet{Becker2022eLife}). The latter finding was found to be tied to the manual lateralization of communicative behaviors (\citet{Becker2022Symmetry}), but not non-communicative actions. Thus the $BaBa21$ template is well-situated to facilitate volumetric tissue-based analyses in a species that is well-poised to elucidate the evolutionary and developmental origins of language, gesture, and brain lateralization.
There is only one longitudinal developmental template for NHPs currently available, for the cynomolgus macaque (\citet{Zhong2022}; \citet{Tan2024}). This species has maturational curves for gray and white matter that peak earlier than in humans, with frontal and temporal lobes showing the largest postnatal increase in volume during the first year with a peak of synapse density at 3 months, after which it declines due to pruning (\citet{Scott2016}). The timing of baboon cohort scanning was based on macaque maturational trajectories, with the first scan timed to occur prior to the gray and white matter maturation, and the second scan at 9 months to capture development after synaptic pruning, during which time a visible inversion of $T2w$ contrast can be observed (\citet{Dehaene2015}). By using both $T1w$ and $T2w$ images in template generation, we took advantage of this additional contrast information for earlier timepoints.
Overall, MRI in nonhuman primates offers numerous advantages for investigating brain development from birth or even prenatal stages. First, most developmental brain MRI databases in humans begin after the age of 4 (e.g., \citet{Raznahan2012}), due to the difficulty of obtaining motion-free images at younger ages. By using anesthetized nonhuman primates, motion-free images may be obtained from any age class from birth onwards. Related to the first point, the ability to use anesthetized subjects relaxes the time constraints required in human studies, permitting a greater number of MR multimodal sequences and better resolution. Third, longitudinal studies in humans pose significant challenges in terms of recruitment and follow-up, which is why most pediatric neuroimaging studies are cross- sectional (e.g., \citet{VanEssen2013}; \citet{Jernigan2016}). With NHPs, access to subjects and control of scan timing can be more precisely controlled. Further, NHP colonies themselves permit better control of environmental variables during development, and the opportunity to introduce behavioral experiments and other interventions.

Finally, the interpolation method proposed in this study enables more targeted morphometric analyses or spatial normalization of cohorts of subjects in all age groups. A template can be generated that precisely fits any age within the template timepoint range. This offers advantages for anatomical analyses; for example, voxel-based morphometry approaches, by reducing the bias that could be introduced from a template generated from a less well-matched period in development. This is particularly relevant for studies examining the period between birth and one year of age, during which $T1w$ and $T2w$ contrasts undergo significant changes (see Supplementary Figure\ref{fig:figure4}).

\subsection{Limitations and Future Directions}

Following a large cohort of NHP over 5 years for repeated acquisitions presents many challenges. From an initial cohort of 33 individuals, at the time of template building, 21 had reached the age to be scanned at timepoint 3. Going forward, as the cohort matures and all subjects’ scans are acquired for timepoint 3, these will be integrated into a future version of
$BaBa21$. Baboons reach sexual maturity at timepoint 3, however, for thoroughness in the developmental cycle, a timepoint 4 will eventually be added when cohort members reach age 7, and subsequently be integrated into the developmental template.
The $BaBa21$ template represents the population as a whole and is not divided according to sex. While this template provides a general reference for the species, and can be used to examine sex differences, future work could develop sex-specific templates and investigate potential effects of sexual dimorphism, which might be especially relevant for timepoints during
and post adolescence.
In this study, we offer a linear interpolation that estimates the curve of development between timepoints, but we acknowledge that cerebral development is non-linear. Thus, the intermediate templates offered here are estimates that will need to be validated and may be improved in the future as actual scans of individuals at these timepoints. By comparing our
intermediate interpolated timepoint templates with an individual scanned at a midpoint between ses-0 and ses-1 (Ozone), we found support for the anatomical accuracy of the volumetric trends, but we acknowledge that our method may not pick up on complex non-linear changes. Nonetheless, these interpolated timepoints are still better targets for registration of scans acquired at intermediate points in development than the original timepoints.
A limitation of the intermediate template generation lies in the interpretation of intensity-based interpolations, particularly in the presence of contrast changes across developmental stages. A “ghosting” effect can be observed in the interpolated images, with the maximum effect potentially occurring between sessions t=0 and t=1 at t=0.5. Further analysis revealed that this effect is not primarily caused by anatomically inconsistent deformations but rather results from contrast inversion in T1-weighted images between consecutive developmental stages. The Supplementary Figure\ref{fig:SupplFig7} illustrates these intermediate deformations and their associated $WM$ surfaces, showing that the interpolated anatomical structures remain spatially coherent despite the apparent intensity artifacts. Importantly, the figure also demonstrates that this effect is negligible between timepoints with similar contrasts (e.g., t=1 and t=2), confirming that the interpolation faithfully preserves anatomical plausibility when image contrast is consistent. These observations indicate that caution is warranted when using an intermediate template (e.g., t=0.5) for studies focusing on groups with similar age or developmental stage, as intensity-related artifacts may influence analyses in cases of large contrast differences. Compared to other acquisitions, timepoint 0 required specific processing interventions. The time of acquisition for this timepoint also had some variation, with scans being taken at 2-4 weeks of age. However, this is still well before the known T1/T2 contrast inversion effects take place (\citet{Young2017};\citet{Cardoso2011}). White matter segmentation was more challenging, not only due to inversion effects, but also the presence of more variable intensities
across white matter. This may be due to different rates of myelination occurring during this early window of brain development. On visual inspection, medial white matter appeared to have greater density than distal white matter, suggesting a directional pattern of myelination.
This variation, in turn, interfered with intensity-based segmentation. This was addressed in $BABACOOL$ by combining $GM$ and $CSF$ segmentation, and subtracting the binary BM, with some manual corrections. In the future, other approaches might be considered, including incorporating three types of priors for white matter at timepoint 0, corresponding respectively to hyper, hypo and “regular” signal intensities. This would also facilitate myelination maturation analysis in early development.

\section{Conclusions}

$BaBa21$ offers a normalization target for researchers working with baboon datasets along the majority of brain development, from neonate to sexual maturity. Additionally, the interpolated intermediate template timepoints generated with $BABACOOL$ permits baboon brains of any age up to 5 years to be registered to the template. TPMs further add to the utility of the
template, allowing for $CSF$ , $WM$, and $GM$ separation for volumetric analyses. Over time, the template will be updated to integrate future timepoints with the goal of capturing the full maturational trajectory of P. anubis.

We anticipate that this novel resource will contribute to comparative primate development work. To facilitate the use of this resource for standardized analyses, cross-study comparisons, and evolutionary-developmental research in the primate neuroimaging community, we have made the multimodal $BaBa21$ template available in the PRIMatE Resource Exchange (PRIME-RE; (\citet{Messinger2021}, \href{https://prime-re.github.io/templates_and_atlases_baboon.html\#Baba21-a-longitudinal-multimodal-template-of-the-developing-baboon-brain}{website}). Comparing baboon, macaque, and human development will permit the elucidation of developmental differences in these species and offer insights into human development and those of model animals used in biomedical research that has implications on human health and cognition.

\section{Data and Code Availability}
The full multimodal $BaBa21$ dataset (BIDS formatted) containing all contrasts and TPMs is
publicly available on the OpenNeuro platform (Dataset ds005424, \citet{LeTroter2025})
Source code and documentation of the pipeline (step by step) for multimodal (4D+t) $BaBa21$ template generation can be found at https://github.com/arnaudletroter/BABACOOL.

\section{Author Contributions}
KLB and ALT wrote the manuscript. YB developed and authored an early version of the manuscript. AM conceived the project. ALT and SAL designed the full processing pipeline. ALT constructed the template and released the code as open source, as well as the final dataset. YB, KKL, and DM contributed to data preprocessing and automated tissue segmentation. SB performed quality control and manual corrections of the segmentations. DM and JS managed and preprocessed the scan data. JS, LR, and AM managed the MR acquisitions. OC and AM supervised the project.

\section{Funding}
This project received funding from the European Research Council under the European Union's Horizon 2020 research and innovation program grant agreement No 716931 - GESTIMAGE - ERC-2016-STG (P.I. Adrien Meguerditchian), from the Agence Nationale de la Recherche (ANR-23-CE28-0029-01, BabOnto  Project, P.I. Adrien Meguerditchian) as well as from grants ANR-16-CONV-0002 (ILCB), and ANR-11- 656 IDEX-0001-02 (A*MIDEX). Y.B. was funded by the Fondation Fyssen. MRI acquisitions were performed at the Center IRM-INT (UMR 7289, AMU-CNRS), a platform member of France Life Imaging network (grant ANR-11-INBS-0006).

\section{Declaration of Competing Interests}
The authors declare no competing interests.

\section{Acknowledgements}
We are very grateful to the Station de Primatologie CNRS, particularly the animal care staff, veterinarians and technicians as well as the administrative staff of the ILCB, the CRPN and the LPC: Nadia Melili, Nadéra Bureau, Frederic Lombardo and Colette Pourpe respectively. 

\bibliographystyle{apalike}
\bibliography{references}

@article{Altmann2010,
  author  = {Altmann, Jeanne and Gesquiere, Laurence and Galbany, Jordi and Onyango, Patrick O. and Alberts, Susan C.},
  title   = {Life History Context of Reproductive Aging in a Wild Primate Model},
  journal = {Annals of the New York Academy of Sciences},
  volume  = {1204},
  pages   = {127--138},
  year    = {2010}
}

@article{Arotcarena2020,
  author  = {Arotcarena, Marie-Laure and Dovero, Sandra and Prigent, Alice and Bourdenx, Mathieu and Camus, Sandrine and Porras, Gregory and Thiolat, Marie-Laure and others},
  title   = {Bidirectional Gut-to-Brain and Brain-to-Gut Propagation of Synucleinopathy in Non-Human Primates},
  journal = {Brain},
  volume  = {143},
  number  = {5},
  pages   = {1462--1475},
  year    = {2020}
}

@article{Avants2008,
  author  = {Avants, Brian B. and Epstein, Charles L. and Grossman, Murray and Gee, James C.},
  title   = {Symmetric Diffeomorphic Image Registration with Cross-Correlation},
  journal = {Medical Image Analysis},
  volume  = {12},
  number  = {1},
  pages   = {26--41},
  year    = {2008}
}

@article{Avants2011,
  author  = {Avants, Brian B. and Tustison, Nicholas J. and Song, Gang and Cook, Paul A. and Klein, Aristeidis and Gee, James C.},
  title   = {A reproducible evaluation of {ANTs} similarity metric performance in brain image registration},
  journal = {NeuroImage},
  volume  = {54},
  number  = {3},
  pages   = {2033--2044},
  year    = {2011},
  month   = {Feb},
  doi     = {10.1016/j.neuroimage.2010.09.025},
  pmid    = {20851191},
  pmcid   = {PMC3065962}
}

@article{Becker2021,
  author  = {Becker, Yannick and Sein, Julien and Velly, Lionel and Giacomino, Laura and Renaud, Luc and Lacoste, Romain and Anton, Jean-Luc and Nazarian, Bruno and Berne, Cammie and Meguerditchian, Adrien},
  title   = {Early Left-Planum Temporale Asymmetry in Newborn Monkeys (Papio anubis)},
  journal = {NeuroImage},
  volume  = {227},
  pages   = {117575},
  year    = {2021}
}

@article{Becker2022eLife,
  author  = {Becker, Yannick and Claidi{\`e}re, Nicolas and Margiotoudi, Konstantina and Marie, Damien and Roth, Muriel and Nazarian, Bruno and Anton, Jean-Luc and Coulon, Olivier and Meguerditchian, Adrien},
  title   = {Broca's Cerebral Asymmetry Reflects Gestural Communication's Lateralisation in Monkeys (Papio anubis)},
  journal = {eLife},
  volume  = {11},
  pages   = {e70521},
  year    = {2022}
}

@article{Becker2022Symmetry,
  author  = {Becker, Yannick and Meguerditchian, Adrien},
  title   = {Structural Brain Asymmetries for Language: A Comparative Approach across Primates},
  journal = {Symmetry},
  volume  = {14},
  number  = {5},
  pages   = {876},
  year    = {2022}
}

@inproceedings{Cardoso2011,
  author    = {Cardoso, M. Jorge and Melbourne, Andrew and Kendall, Giles S. and Modat, Marc and Hagmann, Cornelia F. and Robertson, Nicola J. and Marlow, Neil and Ourselin, Sebastien},
  title     = {Adaptive Neonate Brain Segmentation},
  booktitle = {MICCAI},
  pages     = {378--386},
  year      = {2011}
}

@article{Claidiere2018,
  author  = {Claidi{\`e}re, Nicolas and Amedon, Gameli Kodjo-Kuma and Andr{\'e}, Jean-Baptiste and Kirby, Simon and Smith, Kenny and Sperber, Dan and Fagot, Jo{\"e}l},
  title   = {Convergent Transformation and Selection in Cultural Evolution},
  journal = {Evolution and Human Behavior},
  volume  = {39},
  number  = {2},
  pages   = {191--202},
  year    = {2018}
}

@article{Cusack2018,
  author  = {Cusack, Rhodri and McCuaig, Olivia and Linke, Annika C.},
  title   = {Methodological Challenges in the Comparison of Infant fMRI across Age Groups},
  journal = {Developmental Cognitive Neuroscience},
  volume  = {33},
  pages   = {194--205},
  year    = {2018}
}

@article{Dehaene2015,
  author  = {Dehaene-Lambertz, Ghislaine and Spelke, Elizabeth S.},
  title   = {The Infancy of the Human Brain},
  journal = {Neuron},
  volume  = {88},
  number  = {1},
  pages   = {93--109},
  year    = {2015}
}

@article{Fagot2019,
  author  = {Fagot, Jo{\"e}l and Bo{\"e}, Louis-Jean and Berthomier, Frederic and Claidi{\`e}re, Nicolas and Malassis, Raphaelle and Meguerditchian, Adrien and Rey, Arnaud and Montant, Marie},
  title   = {The Baboon: A Model for the Study of Language Evolution},
  journal = {Journal of Human Evolution},
  volume  = {126},
  pages   = {39--50},
  year    = {2019}
}

@article{Fischer2019,
  author  = {Fischer, Julia and Higham, James P. and Alberts, Susan C. and Barrett, Louise and Beehner, Jacinta C. and Bergman, Thore J. and Carter, Alecia J. and others},
  title   = {Insights into the Evolution of Social Systems and Species from Baboon Studies},
  journal = {eLife},
  volume  = {8},
  doi     = {10.7554/eLife.50989},
  year    = {2019}
}

@article{Friedrich2021,
  author  = {Friedrich, Patrick and Forkel, Stephanie J. and Amiez, C{\'e}line and Balsters, Joshua H. and Coulon, Olivier and Fan, Lingzhong and Goulas, Alexandros and others},
  title   = {Imaging Evolution of the Primate Brain: The next Frontier?},
  journal = {NeuroImage},
  volume  = {228},
  pages   = {117685},
  year    = {2021}
}

@article{Fu2017,
  author  = {Fu, Zhenrong and Lin, Lan and Tian, Miao and Wang, Jingxuan and Zhang, Baiwen and Chu, Pingping and Li, Shaowu and others},
  title   = {Evaluation of Five Diffeomorphic Image Registration Algorithms for Mouse Brain Magnetic Resonance Microscopy},
  journal = {Journal of Microscopy},
  volume  = {268},
  number  = {2},
  pages   = {141--154},
  year    = {2017}
}

@article{Heldstab2018,
  author  = {Heldstab, Sandra A. and Isler, Karin and van Schaik, Carel P.},
  title   = {Hibernation Constrains Brain Size Evolution in Mammals},
  journal = {Journal of Evolutionary Biology},
  volume  = {31},
  number  = {10},
  pages   = {1582--1588},
  year    = {2018}
}

@article{Isler2012,
  author  = {Isler, Karin and van Schaik, Carel P.},
  title   = {Allomaternal Care, Life History and Brain Size Evolution in Mammals},
  journal = {Journal of Human Evolution},
  volume  = {63},
  number  = {1},
  pages   = {52--63},
  year    = {2012}
}

@article{Jernigan2016,
  author  = {Jernigan, Terry L. and Brown, Timothy T. and Bartsch, Hauke and Dale, Anders M.},
  title   = {Toward an Integrative Science of the Developing Human Mind and Brain},
  journal = {Developmental Cognitive Neuroscience},
  volume  = {18},
  pages   = {2--11},
  year    = {2016}
}

@article{Josse2009,
  author  = {Josse, Goulven and Kherif, Ferath and Flandin, Guillaume and Seghier, Mohamed L. and Price, Cathy J.},
  title   = {Predicting Language Lateralization from Gray Matter},
  journal = {Journal of Neuroscience},
  volume  = {29},
  number  = {43},
  pages   = {13516--13523},
  year    = {2009}
}

@article{Kwiecien2014,
  author  = {Kwiecien, Timothy D. and Sy, Christopher and Ding, Yuchuan},
  title   = {Rodent Models of Ischemic Stroke Lack Translational Relevance... Are Baboon Models the Answer?},
  journal = {Neurological Research},
  year    = {2014},
  volume  = {36},
  number  = {5},
  pages   = {417--422}
}

@article{Leigh2004,
  author  = {Leigh, S. R.},
  title   = {Brain Growth, Life History, and Cognition in Primate and Human Evolution},
  journal = {American Journal of Primatology},
  year    = {2004},
  volume  = {62},
  number  = {3},
  pages   = {139--164}
}

@misc{LeTroter2025,
  author  = {Le Troter, Arnaud and Meunier, David and Bryant, Katherine L. and Sein, Julien and Bouziane, Siham and Meguerditchian, Adrien},
  title   = {BaBa21},
  year    = {2025},
  howpublished = {\url{https://doi.org/10.18112/openneuro.ds005424.v1.0.1}}
}

@article{Lizarraga2020,
  author  = {Lizarraga, Stephanny and Daadi, Etienne W. and Roy-Choudhury, Gourav and Daadi, Marcel M.},
  title   = {Age-Related Cognitive Decline in Baboons: Modeling the Prodromal Phase of Alzheimer’s Disease and Related Dementias},
  journal = {Aging},
  year    = {2020},
  volume  = {12},
  number  = {11},
  pages   = {10099--116}
}

@article{Love2016,
  author  = {Love, Scott A. and Marie, Damien and Roth, Muriel and Lacoste, Romain and Nazarian, Bruno and Bertello, Alice and Coulon, Olivier and Anton, Jean-Luc and Meguerditchian, Adrien},
  title   = {The Average Baboon Brain: MRI Templates and Tissue Probability Maps from 89 Individuals},
  journal = {NeuroImage},
  year    = {2016},
  volume  = {132},
  pages   = {526--533},
  month   = {May}
}

@article{Margiotoudi2019,
  author  = {Margiotoudi, Konstantina and Marie, Damien and Claidière, Nicolas and Coulon, Olivier and Roth, Muriel and Nazarian, Bruno and Lacoste, Romain and others},
  title   = {Handedness in Monkeys Reflects Hemispheric Specialization within the Central Sulcus. An in Vivo MRI Study in Right- and Left-Handed Olive Baboons},
  journal = {Cortex},
  year    = {2019},
  volume  = {118},
  pages   = {203--211},
  month   = {September}
}

@article{Marie2018,
  author  = {Marie, Damien and Roth, Muriel and Lacoste, Romain and Nazarian, Bruno and Bertello, Alice and Anton, Jean-Luc and Hopkins, William D. and Margiotoudi, Konstantina and Love, Scott A. and Meguerditchian, Adrien},
  title   = {Left Brain Asymmetry of the Planum Temporale in a Nonhominid Primate: Redefining the Origin of Brain Specialization for Language},
  journal = {Cerebral Cortex},
  year    = {2018},
  volume  = {28},
  number  = {5},
  pages   = {1808--1815}
}

@article{Mars2014,
  author  = {Mars, Rogier B. and Neubert, Franz-Xaver and Verhagen, Lennart and Sallet, Jérôme and Miller, Karla L. and Dunbar, Robin I. M. and Barton, Robert A.},
  title   = {Primate Comparative Neuroscience Using Magnetic Resonance Imaging: Promises and Challenges},
  journal = {Frontiers in Neuroscience},
  year    = {2014},
  volume  = {8},
  pages   = {298},
  month   = {October}
}

@article{Meguerditchian2022,
  author  = {Meguerditchian, Adrien},
  title   = {On the Gestural Origins of Language: What Baboons’ Gestures and Brain Have Told Us after 15 Years of Research},
  journal = {Ethology Ecology \& Evolution},
  year    = {2022},
  volume  = {34},
  number  = {3},
  pages   = {288--302}
}

@article{Meguerditchian2021,
  author  = {Meguerditchian, Adrien and Marie, Damien and Margiotoudi, Konstantina and Roth, Muriel and Nazarian, Bruno and Anton, Jean-Luc and Claidière, Nicolas},
  title   = {Baboons (Papio anubis) Living in Larger Social Groups Have Bigger Brains},
  journal = {Evolution and Human Behavior},
  year    = {2021},
  volume  = {42},
  number  = {1},
  pages   = {30--34}
}

@article{Meguerditchian2006,
  author  = {Meguerditchian, Adrien and Vauclair, Jacques},
  title   = {Baboons Communicate with Their Right Hand},
  journal = {Behavioural Brain Research},
  year    = {2006},
  volume  = {171},
  number  = {1},
  pages   = {170--174}
}

@article{Meguerditchian2011,
  author  = {Meguerditchian, A. and Molesti, Sandra and Vauclair, J.},
  title   = {Right-Handedness Predominance in 162 Baboons (Papio anubis) for Gestural Communication: Consistency across Time and Groups},
  journal = {Behavioral Neuroscience},
  year    = {2011},
  volume  = {125},
  number  = {4},
  pages   = {653--660}
}

@article{Messinger2021,
  author  = {Messinger, Adam and Sirmpilatze, Nikoloz and Heuer, Katja and Loh, Kep Kee and Mars, Rogier B. and Sein, Julien and Xu, Ting and others},
  title   = {A Collaborative Resource Platform for Non-Human Primate Neuroimaging},
  journal = {NeuroImage},
  year    = {2021},
  volume  = {226},
  pages   = {117519}
}

@article{Milham2018,
  author  = {Milham, Michael P. and Ai, Lei and Koo, Bonhwang and Xu, Ting and Amiez, Céline and Balezeau, Fabien and Baxter, Mark G. and others},
  title   = {An Open Resource for Non-Human Primate Imaging},
  journal = {Neuron},
  year    = {2018},
  volume  = {100},
  number  = {1},
  pages   = {61--74.e2}
}

@article{Molesti2016,
  author  = {Molesti, Sandra and Vauclair, Jacques and Meguerditchian, Adrien},
  title   = {Hand Preferences for Unimanual and Bimanual Coordinated Actions in Olive Baboons (Papio anubis): Consistency over Time and across Populations},
  journal = {Journal of Comparative Psychology},
  year    = {2016},
  volume  = {130},
  number  = {4},
  pages   = {341--350}
}

@article{Mugler2000,
  author  = {Mugler, J. P. and Bao, S. and Mulkern, R. V. and Guttmann, C. R. and Robertson, R. L. and Jolesz, F. A. and Brookeman, J. R.},
  title   = {Optimized Single-Slab Three-Dimensional Spin-Echo MR Imaging of the Brain},
  journal = {Radiology},
  year    = {2000},
  volume  = {216},
  number  = {3},
  pages   = {891--899}
}

@article{Mugler1990,
  author  = {Mugler, J. P. and Brookeman, J. R.},
  title   = {Three-Dimensional Magnetization-Prepared Rapid Gradient-Echo Imaging (3D MP RAGE)},
  journal = {Magnetic Resonance in Medicine},
  year    = {1990},
  volume  = {15},
  number  = {1},
  pages   = {152--157}
}

@article{Pozzi2014,
  author  = {Pozzi, Luca and Hodgson, Jason A. and Burrell, Andrew S. and Sterner, Kirstin N. and Raaum, Ryan L. and Disotell, Todd R.},
  title   = {Primate Phylogenetic Relationships and Divergence Dates Inferred from Complete Mitochondrial Genomes},
  journal = {Molecular Phylogenetics and Evolution},
  year    = {2014},
  volume  = {75},
  pages   = {165--183},
  month   = {June}
}

@article{Pujol2002,
  author  = {Pujol, Jesús and López-Sala, Anna and Deus, Joan and Cardoner, Narcís and Sebastián-Gallés, Núria and Conesa, Gerardo and Capdevila, Antoni},
  title   = {The Lateral Asymmetry of the Human Brain Studied by Volumetric Magnetic Resonance Imaging},
  journal = {NeuroImage},
  year    = {2002},
  volume  = {17},
  number  = {2},
  pages   = {670--679}
}

@article{Raznahan2012,
  author  = {Raznahan, Armin and Greenstein, Deanna and Lee, Nancy Raitano and Clasen, Liv S. and Giedd, Jay N.},
  title   = {Prenatal Growth in Humans and Postnatal Brain Maturation into Late Adolescence},
  journal = {PNAS},
  year    = {2012},
  volume  = {109},
  number  = {28},
  pages   = {11366--11371}
}

@article{Richards2016,
  author  = {Richards, John E. and Sanchez, Carmen and Phillips-Meek, Michelle and Xie, Wanze},
  title   = {A Database of Age-Appropriate Average MRI Templates},
  journal = {NeuroImage},
  year    = {2016},
  volume  = {124},
  pages   = {1254--1259}
}

@article{RogersFlattery2020,
  author  = {Rogers Flattery, Christina N. and Rosen, Rebecca F. and Farberg, Aaron S. and Dooyema, Jeromy M. and Hof, Patrick R. and Sherwood, Chet C. and Walker, Lary C. and Preuss, Todd M.},
  title   = {Quantification of Neurons in the Hippocampal Formation of Chimpanzees: Comparison to Rhesus Monkeys and Humans},
  journal = {Brain Structure \& Function},
  year    = {2020},
  volume  = {225},
  number  = {8},
  pages   = {2521--2531}
}

@article{Sawiak2018,
  author  = {Sawiak, S. J. and Shiba, Y. and Oikonomidis, L. and Windle, C. P. and Santangelo, A. M. and Grydeland, H. and Cockcroft, G. and Bullmore, E. T. and Roberts, A. C.},
  title   = {Trajectories and Milestones of Cortical and Subcortical Development of the Marmoset Brain from Infancy to Adulthood},
  journal = {Cerebral Cortex},
  year    = {2018},
  volume  = {28},
  number  = {12},
  pages   = {4440--4453}
}

@article{Scott2016,
  author  = {Scott, Julia A. and Grayson, David and Fletcher, Evan and Lee, Aaron and Bauman, Melissa D. and Schumann, Cynthia M. and Buonocore, Michael H. and Amaral, David G.},
  title   = {Longitudinal Analysis of the Developing Rhesus Monkey Brain Using Magnetic Resonance Imaging: Birth to Adulthood},
  journal = {Brain Structure \& Function},
  year    = {2016},
  volume  = {221},
  number  = {5},
  pages   = {2847--2871}
}

@article{Sun2015,
  author  = {Sun, Xiaofei and Shi, Lin and Luo, Yishan and Yang, Wei and Li, Hongpeng and Liang, Peipeng and Li, Kuncheng and Mok, Vincent C. T. and Chu, Winnie C. W. and Wang, Defeng},
  title   = {Histogram-Based Normalization Technique on Human Brain Magnetic Resonance Images from Different Acquisitions},
  journal = {Biomedical Engineering Online},
  year    = {2015},
  volume  = {14},
  number  = {73}
}

@article{Szabo2012,
  author  = {Szabó, C. Akos and Salinas, Felipe S. and Leland, M. Michelle and Caron, Jean-Louis and Hanes, Martha A. and Knape, Koyle D. and Xie, Dongbin and Williams, Jeff T.},
  title   = {Baboon Model of Generalized Epilepsy: Continuous Intracranial Video-EEG Monitoring with Subdural Electrodes},
  journal = {Epilepsy Research},
  year    = {2012},
  volume  = {101},
  number  = {1-2},
  pages   = {46--55}
}

@article{Tan2024,
  author  = {Tan, Zhiqiang and Nie, Binbin and Wu, Huanhua and Li, Bang and Shang, Jingjie and Zhang, Tianhao and Xiao, Zeyu and others},
  title   = {Brain Development during the Lifespan of Cynomolgus Monkeys},
  journal = {NeuroImage},
  year    = {2024},
  volume  = {120952}
}

@article{VanEssen2013,
  author  = {Van Essen, David C. and Smith, Stephen M. and Barch, Deanna M. and Behrens, Timothy E. J. and Yacoub, Essa and Ugurbil, Kamil and WU-Minn HCP Consortium},
  title   = {The WU-Minn Human Connectome Project: An Overview},
  journal = {NeuroImage},
  year    = {2013},
  volume  = {80},
  pages   = {62--79}
}

@article{Vauclair2005,
  author  = {Vauclair, Jacques and Meguerditchian, Adrien and Hopkins, William D.},
  title   = {Hand Preferences for Unimanual and Coordinated Bimanual Tasks in Baboons (Papio anubis)},
  journal = {Brain Research: Cognitive Brain Research},
  year    = {2005},
  volume  = {25},
  number  = {1},
  pages   = {210--216}
}

@article{VanWoerden2012,
  author  = {van Woerden, Janneke T. and Willems, Erik P. and van Schaik, Carel P. and Isler, Karin},
  title   = {Large Brains Buffer Energetic Effects of Seasonal Habitats in Catarrhine Primates},
  journal = {Evolution},
  year    = {2012},
  volume  = {66},
  number  = {1},
  pages   = {191--199}
}

@article{XinWang2007,
  author  = {Wang, Xin},
  title   = {Laplacian Operator-Based Edge Detectors},
  journal = {IEEE Transactions on Pattern Analysis and Machine Intelligence},
  year    = {2007},
  volume  = {29},
  number  = {5},
  pages   = {886--890}
}

@article{Young2017,
  author  = {Young, Jeffrey T. and Shi, Yundi and Niethammer, Marc and Grauer, Michael and Coe, Christopher L. and Lubach, Gabriele R. and Davis, Bradley and others},
  title   = {The UNC-Wisconsin Rhesus Macaque Neurodevelopment Database: A Structural MRI and DTI Database of Early Postnatal Development},
  journal = {Frontiers in Neuroscience},
  year    = {2017},
  volume  = {11},
  pages   = {29}
}

@article{Zhong2022,
  author  = {Zhong, Tao and Wei, Jingkuan and Wu, Kunhua and Chen, Liangjun and Zhao, Fenqiang and Pei, Yuchen and Wang, Ya and others},
  title   = {Longitudinal Brain Atlases of Early Developing Cynomolgus Macaques from Birth to 48 Months of Age},
  journal = {NeuroImage},
  year    = {2022},
  volume  = {247},
  pages   = {118799}
}

\clearpage 
\section{Supplementary materials}

\begin{figure}[h]
\centering
\includegraphics[width=\textwidth]{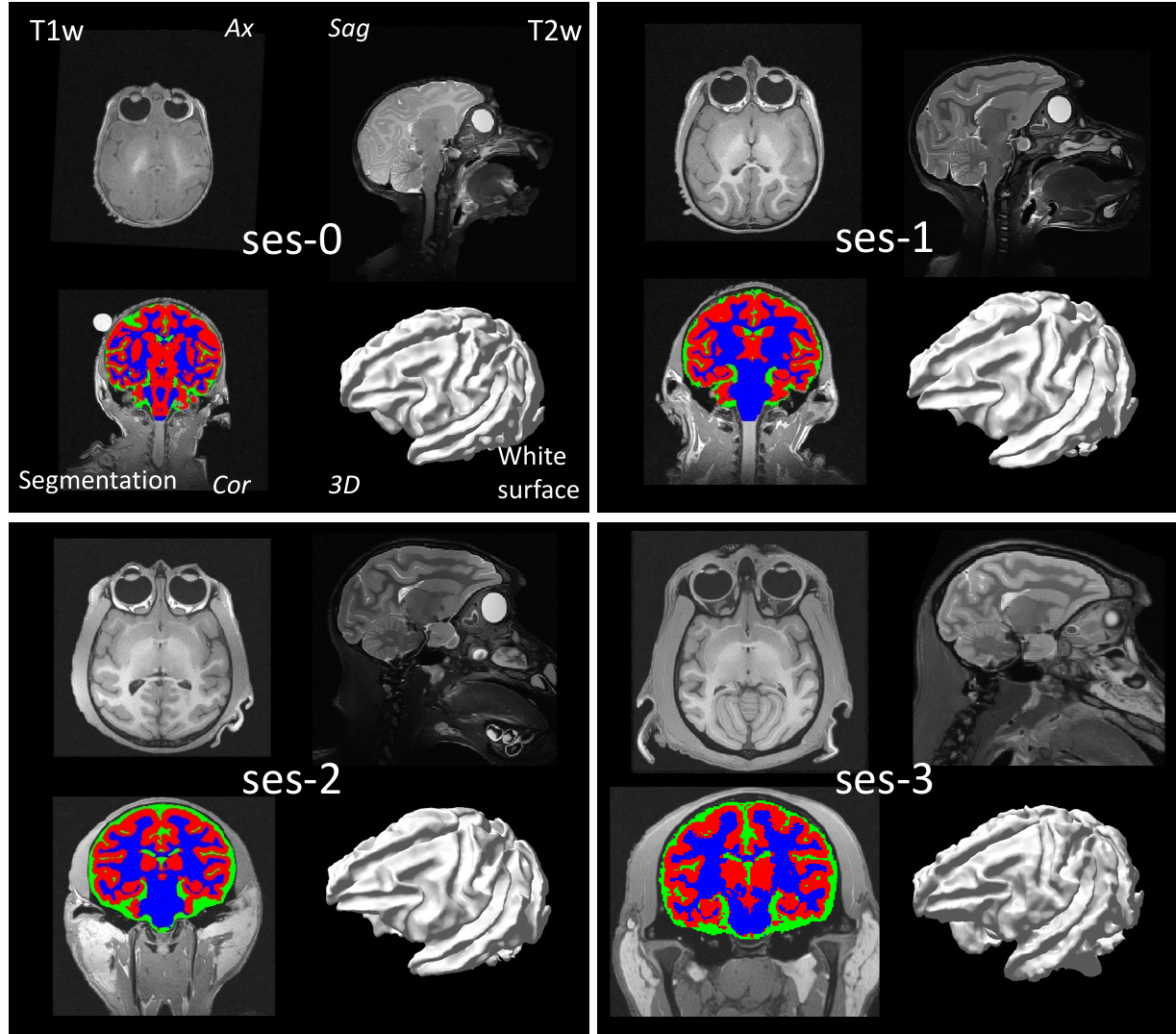}
\caption{(Supplementary) Examples of manual segmentation for a single individual (Prune) across all timepoints; gray matter (green), white matter (red), and $CSF$ (blue).}
\label{fig:SupplFig3}
\end{figure}

\begin{figure}[h]
\centering
\includegraphics[width=\textwidth]{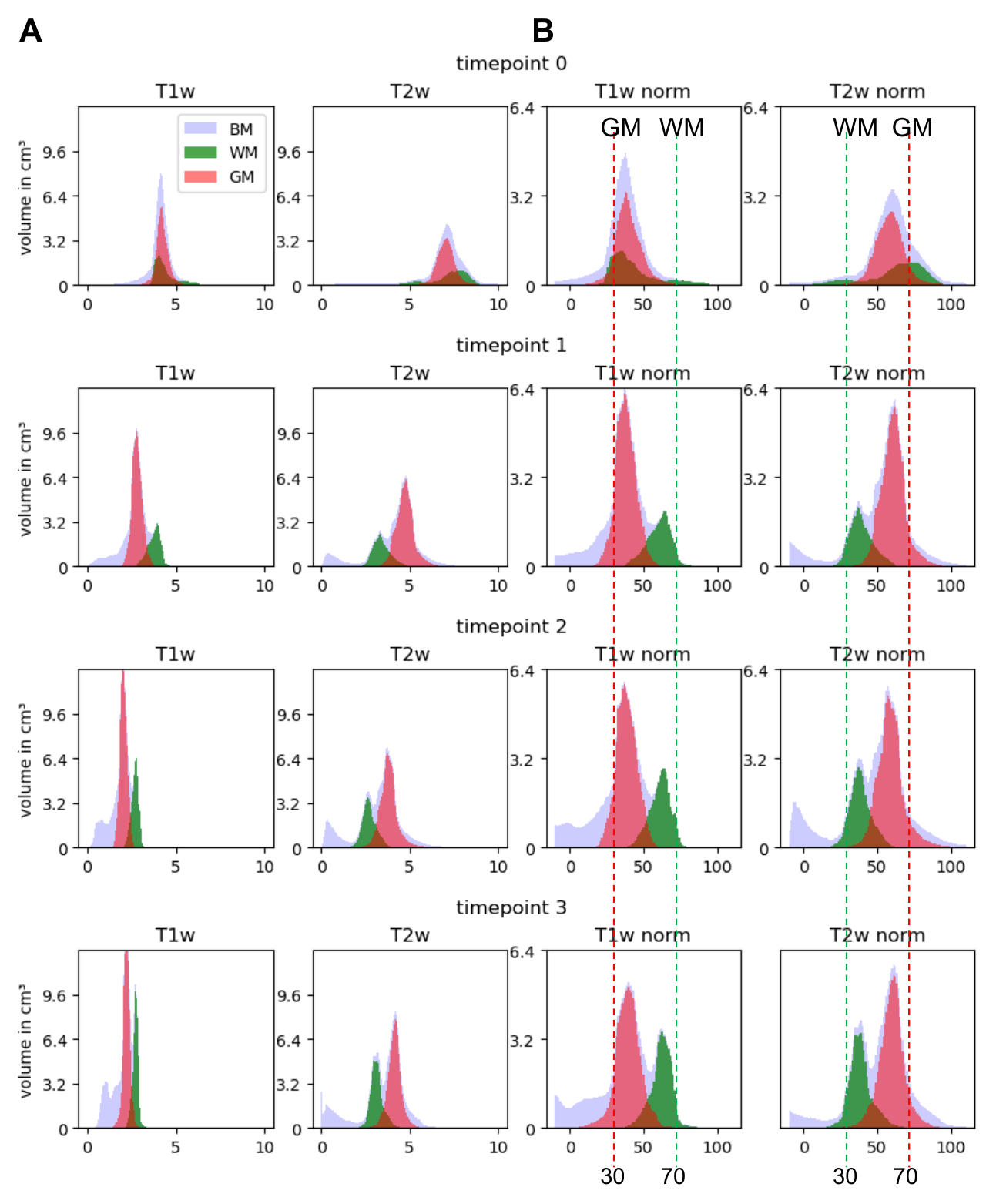}
\caption{(Supplementary) Results of TPM histogram-based normalization for $T1w$ and $T2w$, across all timepoints, before normalization on left (A), after normalization on right (B).}
\label{fig:SupplFig4}
\end{figure}

\begin{figure}[h]
\centering
\includegraphics[width=420pt]{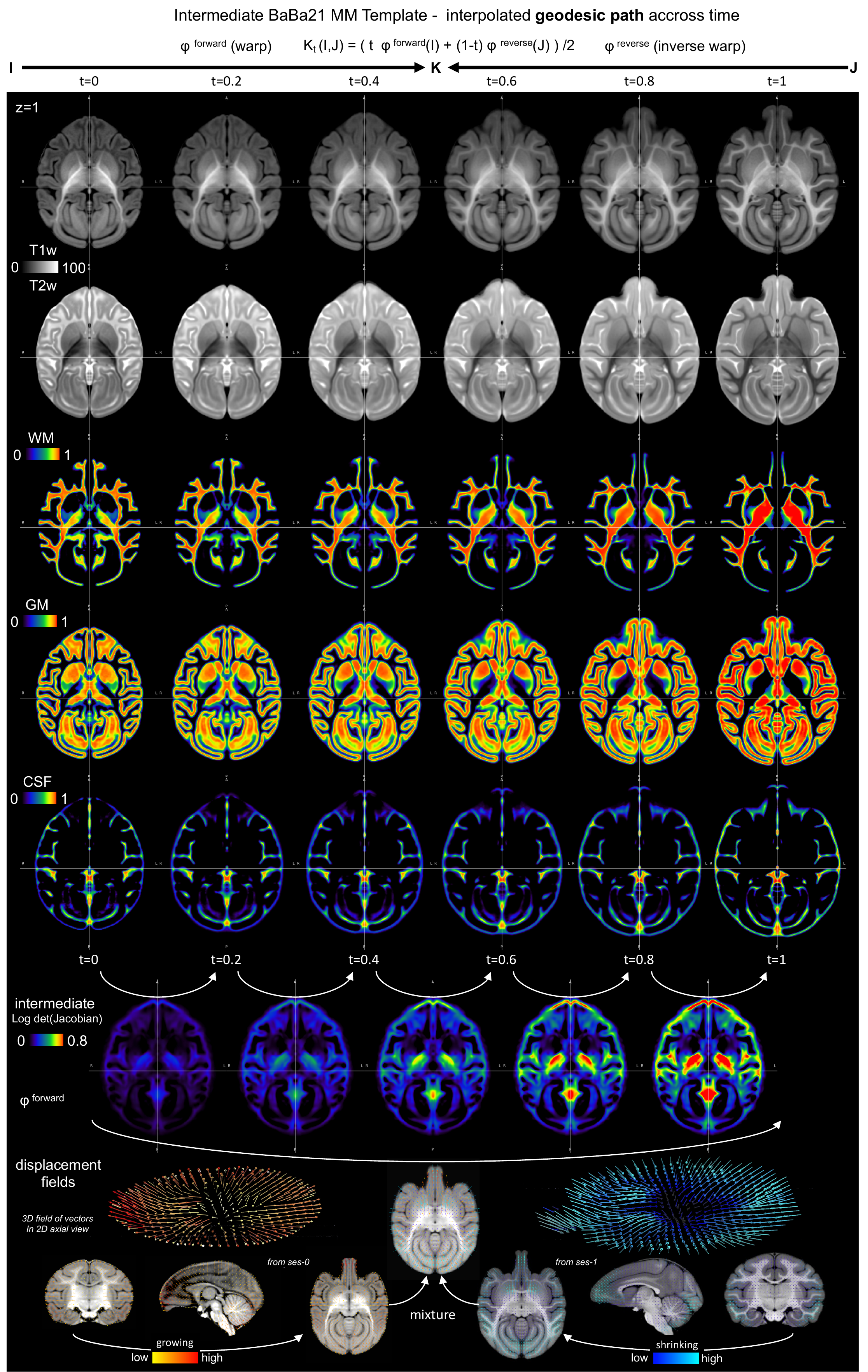}
\caption{(Supplementary) Intermediate $BaBa21$ multimodal template between timepoint 0 and timepoint 1 using interpolated geodesic path across time.}
\label{fig:figure4}
\end{figure}

\begin{figure}[h]
\centering
\includegraphics[width=\textwidth]{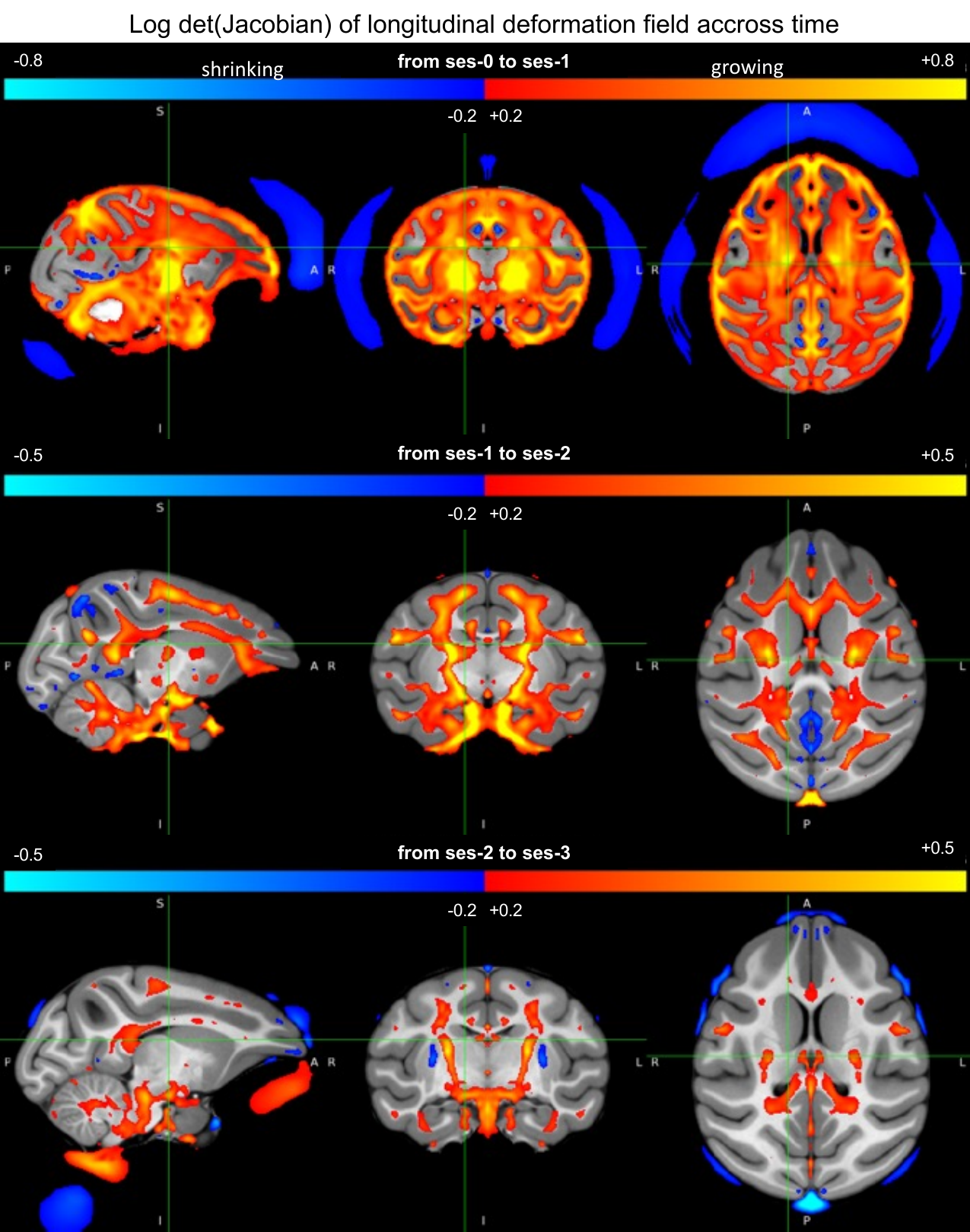}
\caption{(Supplementary) Log Jacobian determinant (log|J|) for the full forward deformation
fields between successive timepoints.
}
\label{fig:SupplFig5}
\end{figure}

\begin{figure}[h]
\centering
\includegraphics[width=\textwidth]{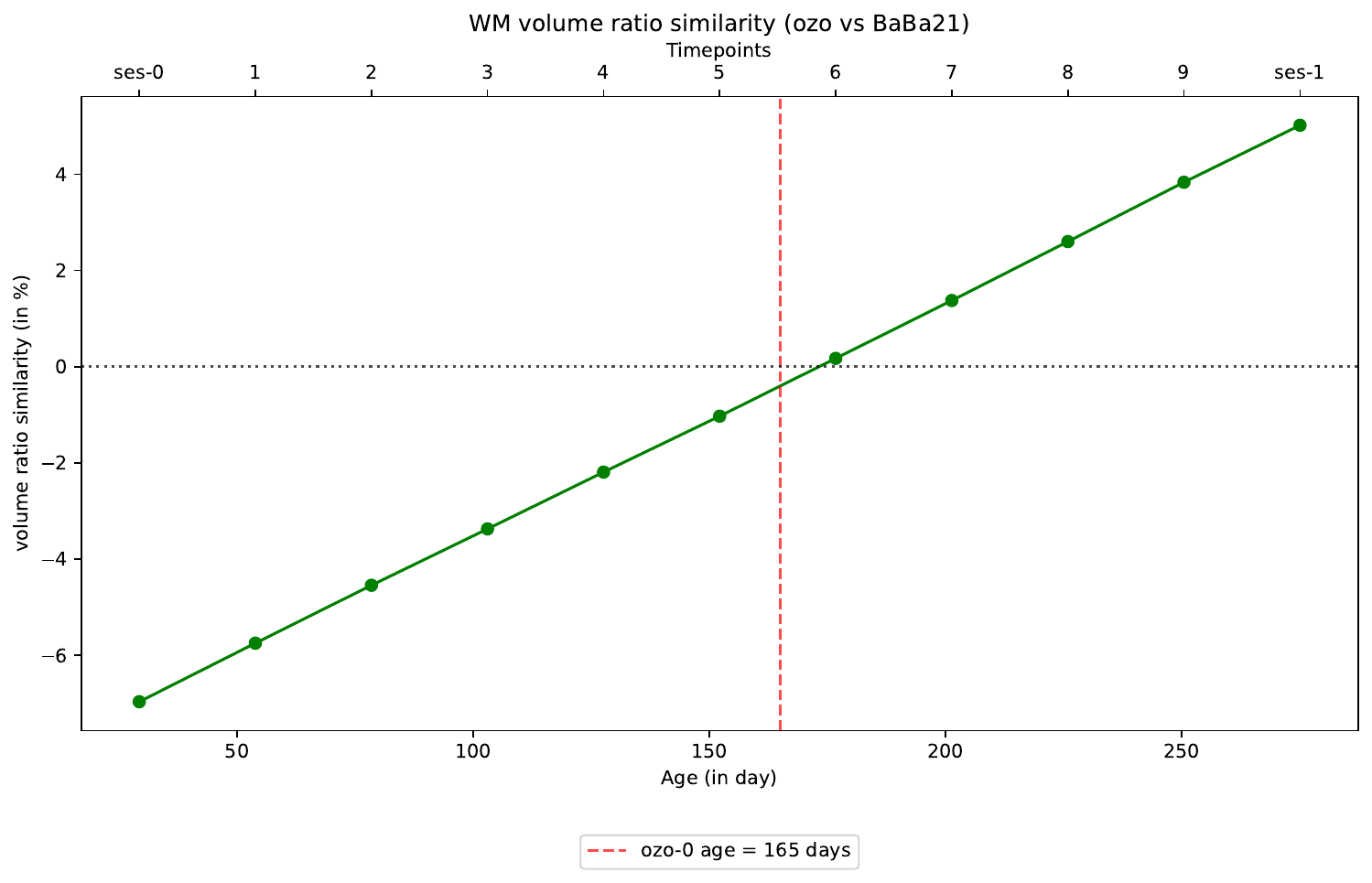}
\caption{(Supplementary) Volume ratio similarity between $Ozone_{ses-0}$ $WM$ volumes (175 days) and
intermediate $BaBa21$ timepoint 6, which corresponds to approximately 175 days..
}
\label{fig:SupplFig6}
\end{figure}

\begin{figure}[h]
\centering
\includegraphics[width=\textwidth]{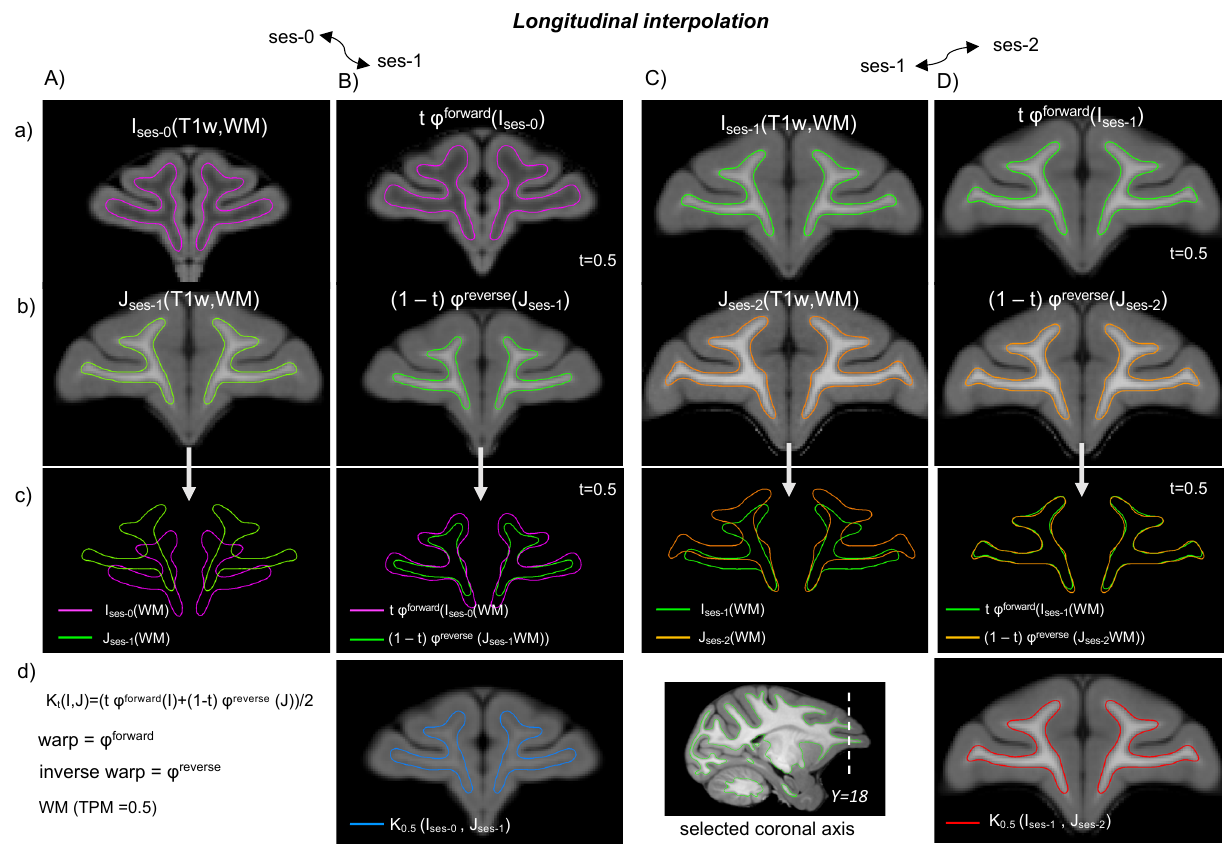}
\caption{(Supplementary)intermediate deformations between developmental sessions.
Columns A and B correspond to interpolations between sessions ses-0 and ses-1, while
columns C and D correspond to interpolations between sessions ses-1 and ses-2.
Row a) shows, in the coronal slice (Y = 18), the $T1w$ volume and white matter (WM) surface
of the source volume I ses-0 along with the forward deformation $\phi$ forward weighted by (t = 0.5).
Row b) displays the target $T1w$ and $WM$ images Jses-1 and the corresponding inverse
deformation $\phi$ reverse (t=0.5)
Row c) presents the superposition of $WM$ surfaces before registration (column A) and after
registration and interpolation (column B)
Row d) shows the results of the interpolation, with $WM$ surfaces in blue highlighting at the
intermediate timepoint (t=0.5)}
\label{fig:SupplFig7}
\end{figure}

\end{document}